\documentclass[fleqn,usenatbib]{mnras}

\usepackage{multirow}
\usepackage{newtxtext,newtxmath}

\usepackage[T1]{fontenc}

\DeclareRobustCommand{\VAN}[3]{#2}
\let\VANthebibliography\thebibliography
\def\thebibliography{\DeclareRobustCommand{\VAN}[3]{##3}\VANthebibliography}

\usepackage{graphicx}	
\usepackage{amsmath}	
\usepackage{subcaption}
\usepackage{url}
\usepackage{comment}

\title[\cmemu]{\cmemu: an emulator of \cmfast\ summary observables}

\author[D. Breitman et al.]{
Daniela Breitman$^{1}$\thanks{E-mail: daniela.breitman@sns.it},
Andrei Mesinger$^{1, 7}$,
Steven G. Murray$^{1,2}$,
David Prelogović$^{1}$,
Yuxiang Qin$^{3,4}$,
Roberto Trotta$^{5,6,7}$
\\
$^{1}$ Scuola Normale Superiore (SNS), Piazza dei Cavalieri 7, Pisa, PI, 56125, Italy\\
$^{2}$ School of Earth and Space Exploration, Arizona State University, Tempe, AZ, USA\\
$^3$ School of Physics, University of Melbourne, Parkville, VIC 3010, Australia\\
$^{4}$ ARC Centre of Excellence for All-Sky Astrophysics in 3 Dimensions (ASTRO 3D)\\
$^{5}$ Scuola Internazionale Superiore di Studi Avanzati (SISSA), Via Bonomea 265, 34136 Trieste, Italy\\
$^{6}$ Imperial Centre for Inference and Cosmology (ICIC), Imperial College, Blackett Laboratory, Prince Consort Road, London SW7 2AZ, U.K.\\
$^{7}$ Centro Nazionale ``High Performance Computer, Big Data and Quantum Computing"
}

\date{Accepted XXX. Received YYY; in original form ZZZ}

\pubyear{2023}

\begin{document}
\newcommand{\note}[1]{{\color{red} #1}}
\newcommand{\tdyn}{t_{\rm dyn}}
\newcommand{\Mhalo}{M_{\rm h}}
\newcommand{\aveMhalo}{\bar{M}_{\rm h}}
\newcommand{\nlea}{\bar{n}_{\rm LAE}}
\newcommand{\zpeak}{z_{\rm peak}}
\newcommand{\taure}{\tau_{\rm reion}}
\newcommand{\tauHII}{\tau_{\rm HII}}
\newcommand{\Dss}{\Delta_{\rm ss}}
\newcommand{\Dmod}{\Delta_{\rm mod}}
\newcommand{\Dextr}{\Delta_{\rm extr}}
\newcommand{\GammaHII}{\Gamma_{\rm HII}}
\newcommand{\aveGammaHII}{\langle \Gamma_{12} \rangle_{\rm HII}}
\newcommand{\avenfHII}{\langle x_{\rm HI} \rangle_{\rm HII}}
\newcommand{\fesc}{f_{\rm esc}}
\newcommand{\QHII}{Q_{\rm HII}}
\newcommand{\qnamesevenone}{ULAS J1120+0641}
\newcommand{\sense}{\textsc{\small 21cmsense}}
\newcommand{\dexm}{\textsc{\small DexM}}
\newcommand{\enzo}{\textsc{\small ENZO}}
\newcommand{\Zsun}{Z_\odot}
\newcommand{\uunit}{{\bf \hat{u}}}
\newcommand{\cmfast}{{\tt 21cmFAST}}
\newcommand{\cmmc}{{\tt 21cmMC}}
\newcommand{\cmemu}{{\tt 21cmEMU}}
\newcommand{\muKK}{\mu {\rm K^2}}
\newcommand{\PkSZ}{[\Delta^{\rm patchy}_{l3000}]^2}
\newcommand{\POV}{[\Delta^{\rm OV}_{l3000}]^2}
\newcommand{\Ptot}{[\Delta_{l3000}]^2}
\newcommand{\reionparams}{\{\zeta, T_{\rm vir}, R_{\rm mfp}\}}
\newcommand{\delz}{\Delta z_{\rm re}}
\newcommand{\zre}{z_{\rm re}}
\newcommand{\HI}{\ion{H}{1}}
\newcommand{\HII}{\ion{H}{2}}
\newcommand{\MgII}{\ion{Mg}{2}}
\newcommand{\OVI}{\ion{O}{6}}
\newcommand{\ef}{x_{e^-}}
\newcommand{\frec}{f_{\rm rec}}
\newcommand{\Ndot}{\dot{N}_{\rm X}}
\newcommand{\Omb}{\Omega_{\rm b}}
\newcommand{\xz}{({\bf x}, z)}
\newcommand{\xzp}{({\bf x}, z')}
\newcommand{\Ts}{T_{\rm S}}
\newcommand{\Tk}{T_{\rm K}}
\newcommand{\aveTs}{\overline{T}_{\rm S}}
\newcommand{\aveTk}{\overline{T}_{\rm K}}
\newcommand{\aveTb}{\overline{T}_{\rm b}}
\newcommand{\denps}{\Delta^2_{\delta \delta}({\bf k}, z)}
\newcommand{\chisq}{\chi^2}
\newcommand{\htwo}{\mathrm{H}_2}
\newcommand{\hone}{\mathrm{H}\textsc{i}}
\newcommand{\hii}{\mathrm{H}\textsc{ii}}
\newcommand{\nf}{\hone}
\newcommand{\Rom}[1]{\uppercase\expandafter{\romannumeral #1}}
\newcommand{\rom}[1]{\lowercase\expandafter{\romannumeral #1}}
\newcommand{\Muv}{M_{\rm 1500}}
\newcommand{\Mvir}{M_{\rm vir}}
\newcommand{\avenf}{\overline{x}_{\hone}}
\newcommand{\nfIGM}{x_{\hone}^{\rm IGM}}
\newcommand{\strom}{HII region}
\newcommand{\lya}{Ly$\alpha$}
\newcommand{\lyb}{Ly$\beta$}
\newcommand{\taudamp}{\tau_{D}}
\newcommand{\taures}{\tau_{R}}
\newcommand{\lobs}{\lambda_{\rm obs}}
\newcommand{\tauobs}{\tau_{\rm obs}}
\newcommand{\tausim}{\tau_{\rm sim}}
\newcommand{\zsource}{z_{\rm s}}
\newcommand{\lcdm}{$\Lambda$CDM}
\newcommand{\fion}{f_{\rm ion}}
\newcommand{\hs}{\hspace{1mm}}
\newcommand{\qnamesfourtwo}{SDSS J1148+5251}
\newcommand{\qnamestwoeight}{SDSS J1030+0524}
\newcommand{\qnamestwotwo}{SDSS J1623+3112}
\newcommand{\qnamestwo}{SDSS J1048+4637}
\newcommand{\qnamefourtwo}{J1148+5251}
\newcommand{\qnametwoeight}{J1030+0524}
\newcommand{\qnametwotwo}{J1623+3112}
\newcommand{\qnametwo}{J1048+4637}
\newcommand{\taub}{\tau_{\rm lim(Ly\beta)}}
\newcommand{\taua}{\tau_{\rm lim(Ly\alpha)}}
\newcommand{\tautot}{\tau_{\rm Ly\alpha}} 
\newcommand{\fobs}{f_\nu^{obs}}
\newcommand{\fcon}{f_\nu^{con}}
\newcommand{\fgamma}{f_\Gamma}
\newcommand{\deltakvec}{\delta({\bf k})}
\newcommand{\deltaxvec}{\delta({\bf x})}
\newcommand{\sigsq}{\sigma^2(M)}
\newcommand{\sig}{\sigma(M)}
\newcommand{\delc}{\delta_c(M,z)}
\newcommand{\Msun}{M_\odot}
\newcommand{\Tvir}{T_{\rm vir}}
\newcommand{\Tvirmin}{T_{\rm vir}^{\rm min}}
\newcommand{\fcoll}{f_{\rm coll}(z, >M_{\rm halo})}
\newcommand{\fcollEPS}{f_{\rm coll}({\bf x}, z, R)}
\newcommand{\Lxsfr}{$\rm{L}_{\rm X}/\rm{SFR}$ }
\newcommand{\deltanlxvec}{\delta_{\rm nl}({\bf x_1}, z)}
\newcommand{\Tcmb}{T_\gamma}
\newcommand{\avedelT}{\delta \overline{T}_b}
\newcommand{\delT}{\delta T_b}
\newcommand{\deldelT}{\delta_{\rm 21}}
\newcommand{\delsq}{\Delta^2_{\rm 21}}
\newcommand{\delNL}{\delta_{\rm nl}}
\newcommand{\Mmin}{M_{\rm min}}
\newcommand{\Mwdm}{m_{\rm wdm}}
\newcommand{\mfp}{R_{\rm mfp}}
\newcommand{\zbegin}{z_{\rm begin}}
\newcommand{\zend}{z_{\rm end}}
\newcommand{\Tgamma}{T_{\gamma, \rm res}}
\newcommand{\avexi}{\bar{\xi}_{12}}
\newcommand{\zon}{z_{\rm on}}
\newcommand{\lmfp}{\lambda_{\rm mfp}}
\newcommand{\lXmfp}{\lambda^{\rm mfp}_{\rm X}}
\newcommand{\toteff}{\epsilon_{\rm fid}}
\newcommand{\taue}{\tau_{\rm e}}
\newcommand{\xhi}{\overline{x}_{\rm HI}}
\newcommand{\Msum}{\mathcal{M}({\bf x}, z)}
\newcommand{\flux}{f({\bf x}, z)}
\newcommand{\Jcrit}{J^{\rm crit}_{21}(M, z)}
\newcommand{\kmps}{\rm km~s^{-1}}
\newcommand{\taueff}{\tau_{\rm eff}^{\rm GP}}
\newcommand\lsim{\mathrel{\rlap{\lower4pt\hbox{\hskip1pt$\sim$}}
        \raise1pt\hbox{$<$}}}
\newcommand\gsim{\mathrel{\rlap{\lower4pt\hbox{\hskip1pt$\sim$}}
        \raise1pt\hbox{$>$}}}
\def\myputfigure#1#2#3#4#5%
{\vskip#5pt\makebox[0pt]{\hskip#2in
\includegraphics[width=#3\textwidth]{#1}}\vskip#4pt\hfill}

\newenvironment{packed_enum}{
\begin{enumerate}
  \setlength{\itemsep}{1pt}
  \setlength{\parskip}{0pt}
  \setlength{\parsep}{0pt}
}{\end{enumerate}}

\newenvironment{packed_item}{
\begin{itemize}
  \setlength{\itemsep}{1pt}
  \setlength{\parskip}{0pt}
  \setlength{\parsep}{0pt}
}{\end{itemize}}

\newcommand{\myurl}{\texttt}
\newcommand{\mytilde}{\texttildelow}



\label{firstpage}
\pagerange{\pageref{firstpage}--\pageref{lastpage}}
\maketitle

\begin{abstract}
Recent years have witnessed rapid progress in observations of the Epoch of Reionization (EoR).  These have enabled high-dimensional inference of galaxy and intergalactic medium (IGM) properties during the first billion years of our Universe.  However, even using efficient, semi-numerical simulations, traditional inference approaches that compute 3D lightcones on-the-fly can take $10^5$ core hours.
Here we present \cmemu: an emulator of several summary observables from the popular \cmfast\ simulation code.  \cmemu\ takes as input nine parameters characterizing EoR galaxies, and outputs the following summary statistics: (i) the IGM mean neutral fraction; (ii) the 21-cm power spectrum; (iii) the mean 21-cm spin temperature; (iv) the sky-averaged (global) 21-cm signal; (v) the ultraviolet (UV) luminosity functions (LFs); and (vi) the Thomson scattering optical depth to the cosmic microwave background (CMB). All observables are predicted with sub-percent median accuracy, with a reduction of the computational cost by a factor of over 10$^4$.
After validating inference results, we showcase a few applications, including: (i) quantifying the relative constraining power of different observational datasets; (ii) seeing how recent claims of a late EoR impact previous inferences; and (iii) forecasting upcoming constraints from the sixth observing season of the Hydrogen Epoch of Reionization Array (HERA) telescope.
\cmemu\ is publicly-available, and is included as an alternative simulator in the public \cmmc\ sampler.
\end{abstract}




\section{Introduction}

The cosmic dawn (CD) of the first luminous objects and eventual reionization of the Universe remain among the greatest mysteries in modern cosmology.
Recent years have seen a dramatic increase in observations of the CD and epoch of reionization (EoR). 
These include: (i) the Lyman-$\alpha$ forest (e.g., \citealt{quasarsFan06,quasarsBecker07, quasarsBecker15,LyaFBosman18,DOdorico23}); (ii) damping wings in quasar spectra (e.g., \citealt{Bolton11,quasarsMortlock11,quasarsBanados18, QDWW20, QDWY20}); (iii) Lyman-$\alpha$ emission from galaxies (e.g., \citealt{LAE10,LAE12, Konno14LAE, Drake17LAEs, Hoag19LAEs, Shibuya19LAEs}); (iv) large-scale polarization of the cosmic microwave background (CMB; e.g., \citealt{Planck18, cmbpol1, cmbpol2}; (v) secondary kinetic Sunaev-Zeldovich (kSZ) CMB anisotropies (e.g., \citealt{Das14ACT,George15SPT,KSZ1}); (vi) upper limits on the cosmic 21-cm power spectrum (PS; e.g., \citealt{PSLOFAR,Trott20PSMWA,HERAobs1, HERAJosh}).
This is set to culminate with a 3D map of H{\footnotesize I} during the first billion years, expected with the upcoming Square Kilometer Array (SKA; e.g., \citealt{SKA, SKA2, AndreiBook}).

In step with these observational advances, Bayesian inference techniques have been developed that allow us to forward model the observations and constrain the parameters of reionization as well as the galaxies responsible (e.g., \citealt{Choudhury05,Mesinger15,21cmMC15, EoRMason18,GreigMWA, LOFARM20, emuPSG20, EoRLyaQin21, LOFAR21,HERAYuxiang,Maity22, Nikolic23}). These rely on efficient simulators, so-called semi-numerical simulations (e.g., \citealt{21cmFASTM07,Thomas09BEARS,Santos10,21cmFASTM11,Visbal12, Ghara15GRIZZLY,21cmFASTv3M20,Trac22AMBER, Maity22,Schneider22}), that typically approximate computationally-expensive radiative transfer (RT) with approaches based on cheap Fast Fourier Transforms (FFTs).  
However, inference can be computationally expensive even with semi-numerical simulations.  
As an example, the recent, state-of-the-art inference using nine galaxy parameters in \citealt{HERAYuxiang} (hereafter HERA22) took $\sim$ 10$^5$ core hours on an HPC center: roughly 10$^5$ likelihood evaluations each taking $\sim$ 1 core hour to simulate the corresponding observables.

A popular alternative approach is to use emulators (e.g., \citealt{emuPSKern17, emuPSSP17, emuShimabukuro17, emuPSJ19, emuPSG20, emuPSM21, emu21cmVAE, Hovav23}). 
Once trained on a set of simulation outputs, an emulator can replace the 
expensive, on-the-fly simulation step in Bayesian inference: a single likelihood evaluation taking $\sim$ 0.1 s instead of $\sim$ 1 hour.  As such, the computational cost is \textit{amortized}, requiring only the initial database of simulations in order to perform subsequent, inexpensive inferences\footnote{Another form of amortized inference is to train neural density estimators to fit the likelihood or likelihood / evidence ratio using simulated data (e.g., \citealt{SBI1, SBI2, SBI3}).  This is referred to as simulation-based inference (SBI), and has the additional benefit of not having to specify an explicit functional form for the likelihood.  SBI has recently been applied to mock 21cm observations \citep{SBI6, SBI5, Prelogovic23, SBI4}, with very promising results.}.  Of course, such amortized inference is restricted to the theoretical model that is used to train the emulator.  Moreover, there is also the additional emulator error to account for, which can be non-negligible for high precision measurements and in corners of parameter space that are poorly sampled (e.g., \citealt{emuPSKern17}, Appendix B of  HERA22). Nevertheless, emulators allow us to rapidly perform many inferences of the same model, testing the impact on the posterior of different likelihood choices, priors, and new data.
Moreover, the emulator error is sub-dominant compared with current, relatively low S/N observations, such as the 21-cm power spectrum upper limits.


Here we present {\tt 21cmEMU}\footnote{\url{https://github.com/21cmfast/21cmEMU}} -- a public emulator of several summary outputs from the semi-numerical code \cmfast\footnote{\url{https://github.com/21cmfast/21cmFAST}}.  These include (i) the volume-averaged hydrogen neutral fraction; (ii) the 21-cm power spectrum; (iii) the global 21cm brightness temperature; (iv) the neutral IGM spin temperature; (v) the ultraviolet (UV) luminosity functions (LFs); (vi) the Thomson scattering optical depth of CMB photons. Our emulator was trained on summary observables from the \textit{withHERA} inference in HERA22, which sampled 
nine astrophysical parameters that characterize galaxy properties.  As a result, our work presents a few important improvements over previous emulators. The unprecedented number of summary outputs allows us to include complementary, multi-wavelength probes of high-$z$ galaxies and the EoR when computing the likelihood.  Moreover, our physically-motivated galaxy parametrization \citep{inputparamsP19} allows us to easily motivate different choices of priors.  We will periodically update {\tt 21cmEMU} to include new summary outputs and astrophysical models.


We showcase our emulator by re-analysing the HERA power spectrum upper limits published in HERA22. We also perform inferences including various combinations of the data, illustrating the constraining power of each probe on the posterior.  One call of  {\tt 21cmEMU} takes $\sim 0.1$ s (compared to $\sim 1$ hour for {\tt 21cmFAST}), with a typical inference finishing in a few hours.

This paper is organized as follows. In Section \ref{sec:data}, we introduce the data used to train the emulator. In Section \ref{sec:emu}, we introduce the network and discuss its architecture, training procedure, and performance. In Section \ref{sec:inference}, we showcase applications of the emulator to EoR/CD inference problems.  We conclude in Section \ref{sec:conclusion}.
We assume a flat $\Lambda$CDM cosmology, with $(\Omega_\Lambda, \Omega_m, \Omega_b,h,\sigma_8,n_s)$ = (0.69, 0.31, 0.049, 0.68, 0.82, 0.97), consistent with results from \citep{Planck18}. Unless stated otherwise, all lengths are in comoving units.

\section{Simulated Dataset}
\label{sec:data}

Our datasets for training and testing are taken from the \textit{withHERA} inference from HERA22, using an increased number of livepoints (18k). This inference used the {\tt Multinest} \citep{MultinestF09} sampler in \cmmc\ \footnote{\url{https://github.com/21cmfast/21CMMC}} \citep{21cmMC15, 21cmMC17, 21cmMC18} with a flat prior on all astrophysical parameters within the ranges shown in all of the corner plots (e.g., Figure \ref{fig:corner_compare}). The likelihood was determined by current observations of the EoR history, galaxy luminosity functions and 21cm upper limits (discussed in detail in Section \ref{sec:21cmFASTinf}). 

We use all of the {\tt Multinest} outputs, including both accepted and rejected samples, resulting in 1.8M parameter samples.  Of these, we randomly select 1.28M for training, 183k for validation and 330k for testing. The database is standardized (subtract the mean and divide by the standard deviation of each summary statistic) before being passed into the network for training.

Our datasets are generated with the public \cmfast\ \texttt{v3} code \citep{21cmFASTM07, 21cmFASTM11, 21cmFASTv3M20}.  \cmfast\ is a semi-numerical simulation code that operates under the assumption that dark matter halos host galaxies which source inhomogeneous, large-scale cosmic radiation fields. Matter density and velocity fields are generated using second-order Lagrangian perturbation theory (e.g., \citealt{2LPT}). Galaxy properties are assigned to dark matter halo fields using empirical scaling relations, following the parametrization in \citet{inputparamsP19}.  The ionizing, X-ray and soft UV cosmic radiation fields sourced by these galaxies are computed with a combination of excursion set and direct integration along the lightcone.  The ionization and thermal state of the IGM gas are then tracked with a set of coupled differential equations, allowing us to compute the various observables discussed below.  The HERA22 runs that form our database assumed a simulation box length of 250 cMpc, with a 128$^3$ grid.  For further details on the simulation code, the interested reader is directed to \citep{21cmFASTM07, 21cmFASTM11, 21cmFASTv3M20}. Below we summarize the astrophysical parameters used as input to \cmemu, and the summary observables that are the corresponding output.

\subsection{Galaxy model and astrophysical parameters}
\label{sec:astro_params}

The input consists of nine parameters that characterize bulk galaxy properties. Two parameters $(f_{\ast, 10}, \alpha_\ast)$ describe the stellar-to-halo mass relation (SHMR), which is a power-law for the faint galaxies (hosted by $M_h \lesssim 10^{12} M_\odot$ halos) that dominate the cosmic radiation fields at $z>5$ (e.g., \citealt{STHMRK12, STHMRD14, STHMRB15, STHMRM16, Sun16UVLF, STHMRY16}):
\begin{equation}
    \frac{M_\ast}{M_h} (M_h) = f_{\ast, 10} 
    \left( \frac{M_h}{M_{10}}\right)^{\alpha_\ast} 
    \left( \frac{\Omega_b}{\Omega_m}\right).
\end{equation}
Here $\Omega_b$ is the universal baryon energy density (as a fraction of the critical energy density), $\Omega_m$ is the total matter (i.e., cold dark matter and baryon) energy density, and $f_\ast \equiv f_{\ast, 10} \left( \frac{M_h}{M_{10}}\right)^{\alpha_\ast}  \in [0,1]$ is the stellar fraction, with $f_{\ast, 10}$ corresponding to the fraction of galactic gas in stars normalized to the amount in a halo of mass $M_{10} \equiv 10^{10} M_\odot$, and $\alpha_\ast$ the power-law index.

Star formation is assumed to occur on a time-scale that goes with the Hubble time, $H^{-1}(z)$, (or analogously the dynamical time, which also scales with the Hubble time during matter domination):
\begin{equation}
    \Dot{M}_\ast = \frac{M_\ast}{t_\ast H^{-1}(z)}.
\end{equation}
The characteristic star formation timescale, $t_\ast \in [0,1]$, is another free parameter.

The typical ionizing escape fraction, $\fesc (M_h)\in [0,1]$ is similarly described by a power-law (e.g., \citealt{Paardekooper15PL,Kimm17PL, Lewis20PL}):
\begin{equation}
    \fesc (M_h) = f_{\rm esc, 10} \left( \frac{M_h}{M_{10}} \right)^{\alpha_{\rm esc}},
\end{equation}
with two free parameters: the normalization, $f_{\rm esc, 10}$, and the power-law index, $\alpha_{\rm esc}$.

Star formation is suppressed in small mass halos due to inefficient gas cooling and/or  feedback (e.g., \citealt{SFQH97, SFQS03, SFQO08, SFQS13, SFQX16, SFQO20, SFQM20}).  We account for this suppression by assuming only a fraction $\exp \left( -M_{\rm turn} / M_h \right)$ of halos host active star forming galaxies.  The characteristic halo mass scale below which the abundance of galaxies is exponentially suppressed, $M_{\rm turn}$, is another free parameter.

The specific X-ray luminosity escaping the galaxies is also taken to be a power-law in energy (e.g., \citealt{Das17}), $L_X \propto E^{-\alpha_X}$, with the index $\alpha_X$ left as a free parameter.  The luminosity is normalized via the soft band X-ray luminosity per unit SFR, another free parameter:
\begin{equation}
    L_{X < 2\text{keV}} / \text{SFR} = \int_{E_0}^{2\text{keV}} \text{dE} \ L_X / \text{SFR},
\end{equation}
where $E_0$, the last input parameter, is the minimum energy of X-ray photons capable of escaping their host galaxy.

In summary, the nine input parameters are:
\begin{enumerate}
\item $\boldsymbol{f_{\ast, 10}}$: normalization of the SHMR, defined at $M_h = M_{10}$.
\item $\boldsymbol{\alpha_\ast}$: power-law index of the SHMR.
\item $\boldsymbol{f_{\rm esc, 10} }$: normalization of the ionizing escape fraction to halo mass relation, defined at $M_h = M_{10}$.
\item $\boldsymbol{\alpha_{\rm esc}}$: power-law index of the ionizing UV escape fraction to halo mass relation.
\item $\boldsymbol{t_\ast}$: characteristic star formation timescale, defined as a fraction of the Hubble time.
\item $\boldsymbol{M_{\rm turn}/{\rm M}_{\odot}}$: characteristic mass below which halos become exponentially less likely to host an active star forming galaxy.
\item $ \boldsymbol{\frac{L_{\rm X<2keV}/{\rm SFR}}{{\rm erg\ s^{-1}\ M_{\odot}^{-1}\ yr}}}$: soft-band X-ray luminosity per unit SFR escaping the galaxies.
\item $\boldsymbol{E_0/{\rm keV}}$: minimum X-ray energy that can escape the galaxies.
\item $\boldsymbol{\alpha_{\rm X}}$: power-law index of the X-ray spectral energy distribution (SED).
\end{enumerate}
This simple parametrization is easy to interpret physically and is consistent with observations of the UV LFs as well as the scaling relations found in galaxy simulations and semi-analytic models.

\subsection{Observational summaries}

For a given set of cosmological and astrophysical parameters, \cmfast\ calculates the corresponding 3D lightcones of IGM properties.  When performing inference, these lightcones are generally compressed into summary statistics that are compared directly with observations.  Here we do not attempt to directly emulate the 3D lightcones of the various cosmological quantities, and instead only emulate the following summary observables (motivated by existing EoR/CD observations discussed in Section \ref{sec:21cmFASTinf}):
\begin{enumerate}
\item $\boldsymbol{\xhi(z)}$ --- the volume-averaged neutral fraction of hydrogen and helium as a function of redshift (also commonly referred to as the EoR history).
\item $\boldsymbol{\aveTb(z)}$ --- the volume-averaged (global) 21cm brightness temperature (e.g., \citealt{Madau97, Furlanetto06, Pritchard12}): 
\begin{align}
\label{eq:brightness_temperature}
    T_{\rm b}(\boldsymbol{x}, z) &=  \frac{T_{\rm S} - T_{\rm R}}{1+z}(1-e^{\tau_{21}})\\ 
\nonumber &\approx  27 \ x_{\rm HI} (1 + \delta_b) \left( \frac{\Omega_b h^2}{0.023}\right) \left( \frac{0.15}{\Omega_m h^2} \frac{1+z}{10}\right)^{1/2} \ \rm{mK} \\
\nonumber &\times \left( \frac{T_{\rm S} - T_{\rm R}}{T_{\rm S}}\right) \left[ \frac{\partial_r v_r} {(1+z)H(z)}\right],
\end{align}
where $\tau_{21}$ is the 21cm optical depth of the intervening gas, $\delta_b \equiv \rho/\bar{\rho} - 1$ is the baryon overdensity, with $\rho$ being the baryon density, and $T_{\rm S}$ and $T_{\rm R}$ are the spin and background temperatures, respectively.  We assume throughout that the radio background is provided by the CMB, $T_{\rm R} = T_{\rm CMB}$ is the temperature of the CMB.  We note that \cmfast\ computes the brightness temperature at each cell location, ${\bf x}$, using the exact expression in the first line of the equation above; the second line is a Taylor expansion in the limit of $\tau_{21} \ll 1$ that provides physical intuition.
\item $\boldsymbol{\aveTs(z)}$ --- the mean spin temperature of the neutral IGM as a function of redshift.  The IGM spin temperature is only defined for neutral hydrogen that is outside of the cosmic HII regions that surround galaxies.  Specifically, the volume average is performed over those cells in the simulation box with $\xhi \geq$ 95 \%.
\item $\boldsymbol{\delsq(k, z)}$ --- spherically-averaged 21cm power spectrum (PS): $\Delta^2_{21}(k,z)\left[\rm{mK}^2\right] \equiv k^3/(2 \pi^2) \langle \tilde{T}_{b} \tilde{T}_{b}^\ast\rangle$, where $k = |\boldsymbol{k}|$, and $\tilde{T}_{b}(\boldsymbol{k}, z)$ is the Fourier dual of the brightness temperature from eq. (\ref{eq:brightness_temperature}).
\item $\boldsymbol{\phi(\Muv, z)}$ --- the non-ionizing UV luminosity function (UV LF), defined as the number density of galaxies per UV magnitude, $\Muv$, as a function of redshift.  The $\sim$1500 \AA\ rest frame luminosity is calculated from the SFR:
$\dot{M}_\ast (M_h,z) = \mathcal{K}_{\rm UV} \times L_{\rm UV},$ where  $\mathcal{K}_{\rm UV} = 1.15 \cdot 10^{-28} \rm{M}_\odot \rm{ yr}^{-1} \ \rm{Hz} \ \rm{s}\ {erg}^{-1}$ assumes a Salpeter initial mass function (e.g., \citealt{MadauD14, Sun16UVLF}).
The UV luminosity is related to the AB magnitude using \citep{Oke83UVLF}: $\log\left( \frac{L_{\rm UV}}{\rm{erg} \ {s}^{-1} \ {Hz}^{-1}}\right) = 0.4 \times (51.63 - M_{\rm UV}).$
\item $\boldsymbol{\taue }$ --- the Thompson optical depth to the last scattering surface (LSS):
$\tau_e = \sigma_T \int_0^{z_{\rm{LSS}}} dz \ \vline \frac{c dt}{dz} \vline \ n_e $,
where $\sigma_T$ is the Thompson scattering cross section and $n_e$ is the electron number density calculated assuming hydrogen and helium are singly ionized at a fraction $(1-\avenf)$ and that helium is doubly ionized at $z < 3$.
\end{enumerate}
Although the last two quantities are computed analytically by \cmfast\, and are therefore reasonably fast, we still emulate them for two reasons.  The first is to provide users of \cmemu\ with a standalone package. The second is that the analytic calculation is still slower than the emulator prediction time: emulation reduces the runtime from $\sim 1$ s to $\lesssim$ 50 ms for a single parameter combination ($\lesssim$ 1 ms per parameter set if in a large ($\gtrsim 100$) batch), with a relatively low emulation error (see Section \ref{sec:emu}).

We use 84 redshift bins in the range $z \sim 5 - 35$ for all summaries except the 21cm PS. 
For the 21cm PS we exclude high redshift bins that generally have a very weak signal, keeping 60 redshift bins spanning $z \sim 6 - 21$, and 12 $k$ bins spanning $k \sim 0.04 - 1$ Mpc$^{-1}$.  We also floor the PS values to 0.1 mK$^2$, in order to reduce the dynamic range of the data and improve training. We note that the value of the floor is an order of magnitude smaller than the accuracy of the \cmfast\ simulator itself (e.g., \citealt{21cmFASTM11,Zahn11}), and thus has no effective impact on the accuracy of our emulator.
\section{Emulator architecture and performance}
\label{sec:emu}

\begin{figure*}
    \centering
    \includegraphics[width = 0.8\textwidth]{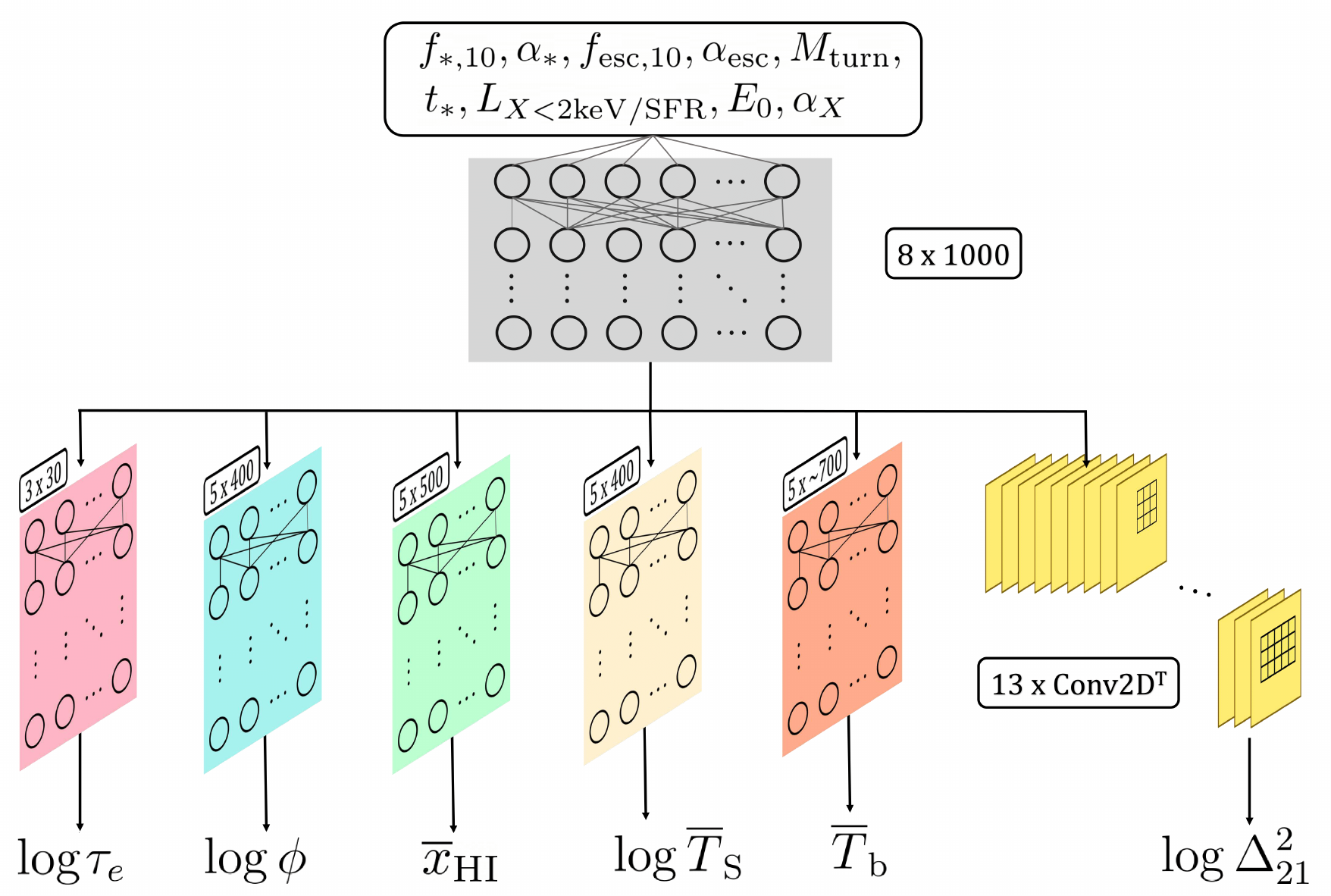}
    \caption{Schematic of the \cmemu\ architecture. Astrophysical parameters ({\it top}; c.f. Section \ref{sec:astro_params} are inputted through a large block of fully-connected layers. The output from this shared block is then passed on into five blocks (much smaller than the shared block). The pink, blue, green, yellow, and orange fully-connected branches output the Thomson scattering optical depth, UV LFs, mean hydrogen neutral fraction, spin temperature, and global signal, respectively. The output from the shared block is also reshaped into an image and is passed into a 2D convolutional neutral network which outputs the 21cm power spectrum. The convolutions gradually build the PS image. The window size varies among the layers. The number of filters (yellow layers) decreases toward the end of the CNN.}
    \label{fig:network_diagram}
\end{figure*}

 \cmemu\ is implemented using {\tt Tensorflow} \citep{tensorflow2015-whitepaper, tensorflow_developers_2022_6574269}, with an architecture consisting of (see diagram in Fig. \ref{fig:network_diagram}):
\begin{itemize}
    \item one large block (8 layers with 1k nodes each) of fully-connected (dense) layers whose output is fed into all of the branches.
    \item one branch per summary observable.
\end{itemize}

\begin{figure}
    \centering
    \includegraphics[width = 0.45\textwidth]{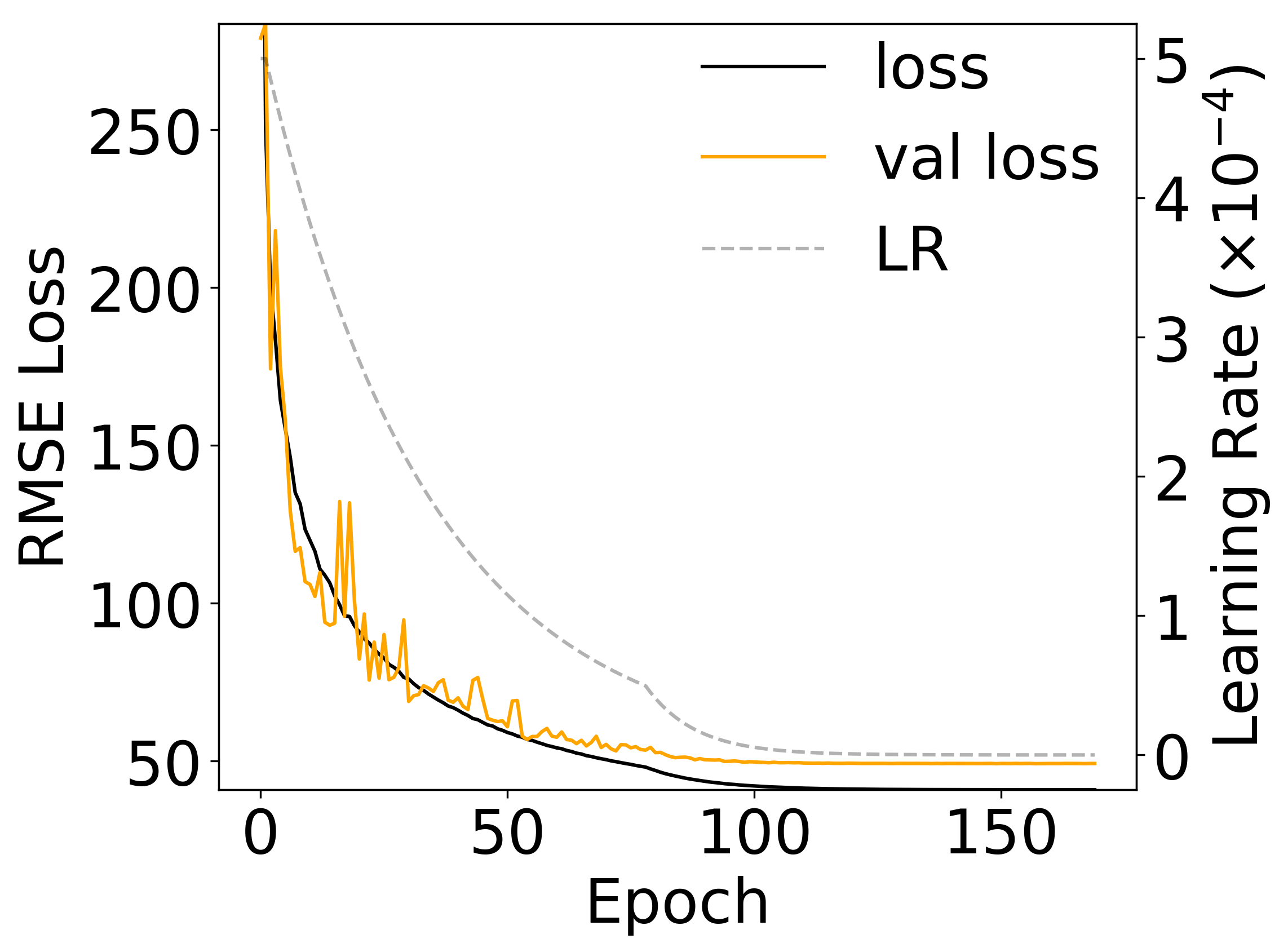}
    \caption{Training loss (black line) and validation loss (orange line) as a function of training epoch. The learning rate curve is also shown with the dashed gray line and the corresponding right axis.}
    \label{fig:training_loss}
\end{figure}

Since the 21-cm power spectrum is a smooth function of wavemode and redshift (e.g., Fig. \ref{fig:PS_compare}), it can be interpreted as a 2D image.  Therefore we use a convolutional neural network (CNN) in the 21-cm PS branch and fully-connected layers in the other branches.  
Note that the branches are not connected with one another. The only nodes they have in common are those from the main block which each branch receives as input.

The network trains on all of the summaries at once (i.e., multi-task learning), using a weighted sum of root mean squared error (RMSE) losses with one loss term per branch, where each branch loss has a different weight. We assign the largest weight to the 21-cm PS branch as it is higher dimensional with the largest dynamic range, and thus more difficult to learn. The final set of weights chosen is obtained from a trial of about 50 different weight combinations with the goal of choosing the best weights such that the 21-cm power spectrum, brightness temperature, and neutral fraction are learned best. The performance of the other summaries is not significantly affected by the choice of weights. These trials are done ad hoc since the training is computationally expensive.

We perform a few tests to motivate the importance of the block of fully-connected layers. First, we train a network equivalent to the brightness temperature branch alone i.e., whose input is the astrophysical parameters that are straight away passed into the brightness temperature branch of fully-connected layers. We find that the median brightness temperature fractional error over the test set in this network  is $\sim$ 45\% larger than the one in the final network. \footnote{We also test slightly changing the brightness temperature branch itself: adding an additional layer and increasing the number of nodes increases the median fractional error (see below for specific definition) by about 50\%, while increasing the number of layer nodes slightly and adding one additional layer increases it by about 20\%.}
This means that, on average\footnote{We did not perform this test for all of the other summaries. We did perform it for the 21-cm power spectrum and found that the performance of the final network is a few percent better than that of the CNN branch alone.}, our final architecture performs better than just having individual networks for each summary. The final architecture can contribute to improving the performance in two ways: (i) combining the losses of the summaries allows the network to learn from the correlations between the summaries; and (ii) simply making the network larger and deeper. To test the relative importance of (i) and (ii),
we train a network without the shared block but with the rest of the architecture the same. This significantly reduces the number of trainable parameters in the network (by about 50\%), but still allows different summaries to influence each other through the shared loss. We do see an increase of up to a few percent in the median and 68\% CL of the fractional error for the smaller network as expected. Most notably, for the brightness temperature we see an increase of $\sim$ 1\%,  $\sim$13\%, and $\sim$27\% for the median, 68\% CL, and 95\% CL of the fractional error, respectively. We conclude that combining the losses of all the summaries is the main cause of performance improvement, while the large shared block is needed to get the best performance for the most challenging summaries: the brightness temperature and 21-cm power spectrum.

In Figure \ref{fig:training_loss}, we show the total training and validation losses as a function of epoch in black and orange, respectively.   We also show the learning rate schedule used during training with a dashed gray line.  We see a smooth decline of the validation loss up to $\sim100$ epochs.  Our final network is taken at the minimum of the validation loss, at epoch 150.  The training takes about eleven GPU hours ($\sim$3.5 min per epoch) with the full database (1.8M samples).


Below, we discuss the branch architecture and performance for each summary observable in turn, summarizing the results in Table \ref{tab:network_summary}.  Throughout, we illustrate the emulator performance using examples from the test set, as well as the distributions of absolute differences (Abs Diff) and fractional errors (FE) over the entire test set.  The latter two are defined for each observational summary, $y$, as:
\begin{align}\label{eq:FE}
    \rm{Abs~Diff} &\equiv | y_{\rm true} - y_{\rm pred}| \\
    \rm{FE (\%)} &\equiv \frac{ \rm{Abs~Diff} }{\max{(|y_{\rm true}|, y_{\rm floor}})} ,
\end{align}
where $y_{\rm true}$ refers to the \cmfast\ direct simulation output and $y_{\rm pred}$ is the corresponding \cmemu\ prediction. We compute the above averaged over different bins in $y$ and/or different models in the test set, as described below.
One drawback of the FE metric is that it can diverge to infinity as the denominator goes to zero.  To avoid this, we use floors for the values of the denominator: $\log(\Delta^2_{21,\rm floor}) = 0.1$; $\bar{T}_{b, \rm floor} = 5$ mK, and $\bar{x}_{\rm{HI}, \rm{floor}} = 10^{-4}$. The specific values of these floors was chosen relatively arbitrarily; however, they are lower than the expected accuracy achievable by any near term experiment.\footnote{For the 21-cm power spectrum for example, the expected mean noise level from thermal noise and sample variance for a 1000hr observation with the SKA1-low instrument is $\gtrsim$0.1 mK$^2$ (e.g., see Figure 2, bottom left panel in \citealt{Kaur20}). Similarly, global signal experiments have measurement noise that is orders of magnitude larger than the floor value we chose (e.g. \citealt{Singh22, Murray22}), and are instead limited mostly by foregrounds and instrument systematics. For the mean neutral fraction, estimates have typical uncertainties of order 0.1 (see e.g. \citealt{Greig22} and references therein),  orders of magnitude larger than the floor value we use.} The other summaries, $\taue, \aveTs$, and UV LFs do not have a floor value.

\subsection{The 21-cm power spectrum} \label{sec:ps}

The power spectrum branch consists of thirteen 2D convolution layers with wide (up to 7 redshift bins $\times$ 3 k bins) kernels and two upsampling layers that gradually build the ($k, z$) PS image based on the output of the shared block, as seen in Figure \ref{fig:network_diagram}. We use a pixel-based RMSE loss, weighted by the inverse of the estimated thermal noise corresponding to a 1000h SKA1-low observation (taken from \citealt{David22}; for more details see Section 2.2.1 in that work).  Weighting by the inverse of the noise forces the CNN to be more accurate in ($k, z$) bins that are easier to observe: generally corresponding to lower redshifts and larger scales.

\begin{figure*}
    \centering
    \includegraphics[width = \textwidth]{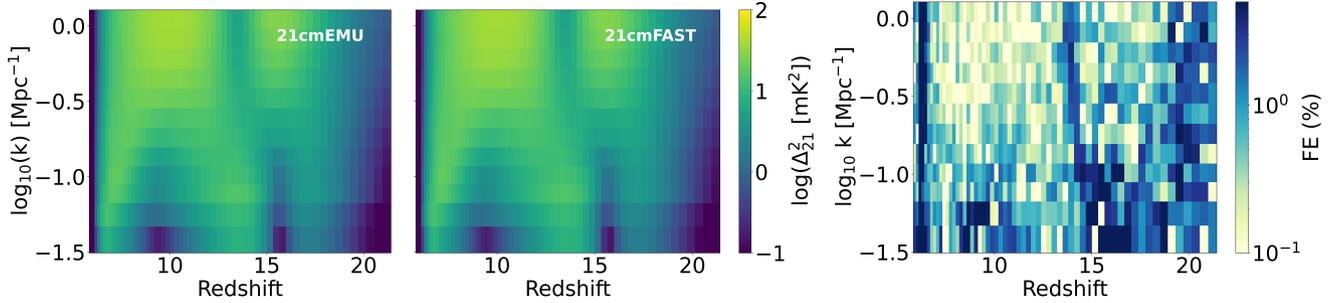}
   \caption{The spherically-averaged 21-cm power spectrum as a function of wavemode and redshift for a sample in the test set.  The \cmemu\ prediction is shown on the left while the \cmfast\ result is on the right.  This sample has a 21-cm PS FE that is roughly comparable to the median value of the whole the test set, and can thus be considered representative of the emulator performance. The rightmost panel shows the fractional error for this single sample.}
   \label{fig:PS_compare}
\end{figure*}

In Figure \ref{fig:PS_compare}, we compare the emulator prediction for the 21-cm power spectrum with its corresponding target from \cmfast. We show a single sample from the test set, with the \cmemu\ prediction on the left and the \cmfast\ target in the middle panel.  This sample was chosen as it has the closest median fractional error to that of the entire test set; thus it can be considered representative of the typical emulator performance.  It is difficult to see a difference between the two PS with the naked eye.  We the FE of this single sample in the rightmost panel.  The FE is generally sub-percent, rising to $\sim$percent in regions of low power.

In these 2D images we clearly see the well-known trend of three peaks in the redshift evolution of the large-scale 21-cm PS and two peaks in the small-scale evolution (e.g., \citealt{Pritchard07}).  In general, the features evolve smoothly over $(k, z)$, showcasing why we use a CNN in the 21-cm PS branch of \cmemu.

\begin{figure*}
\begin{subfigure}{\linewidth}
    \includegraphics[width = 0.45\columnwidth]{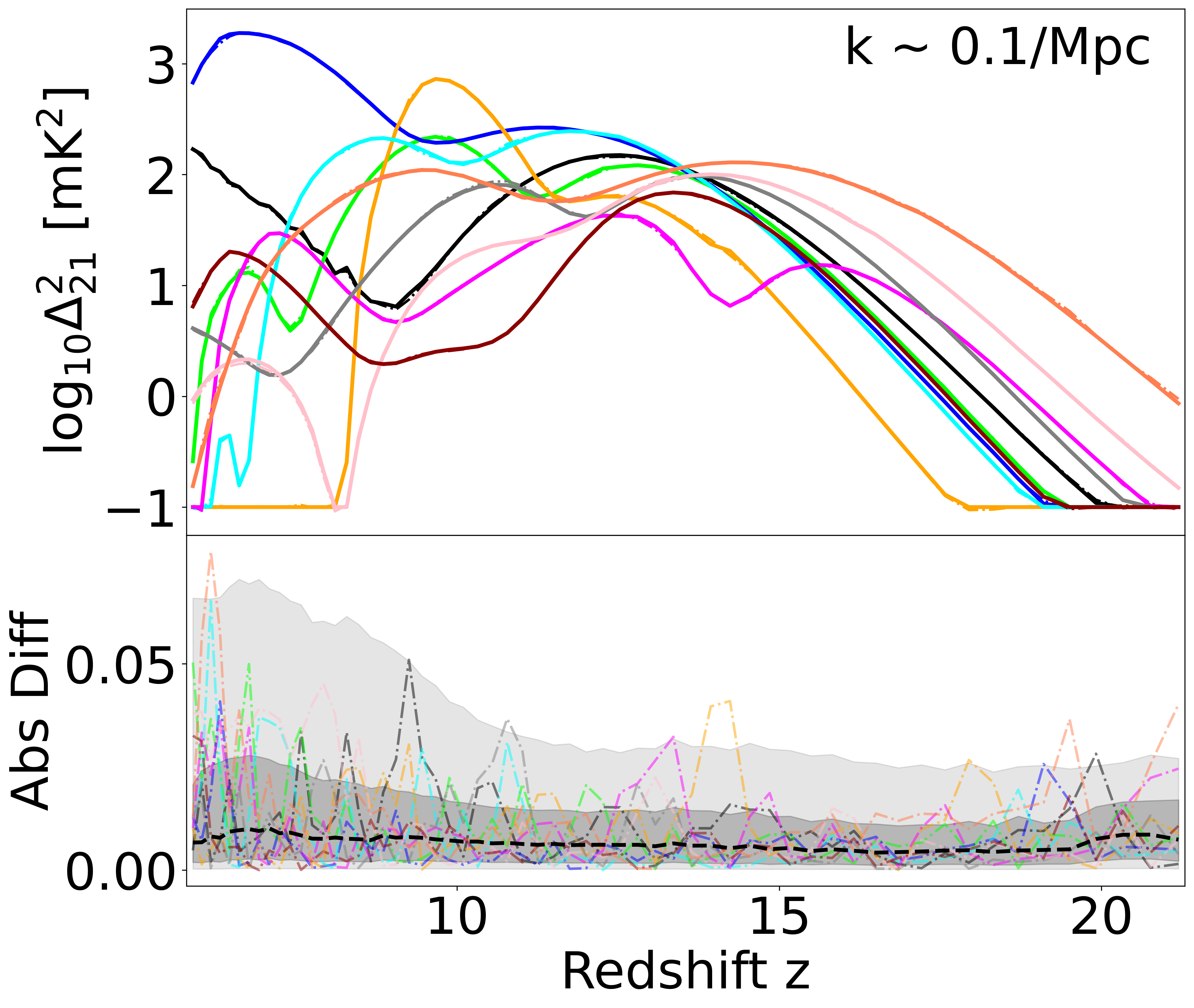} \hfill
    \includegraphics[width = 0.45\columnwidth]{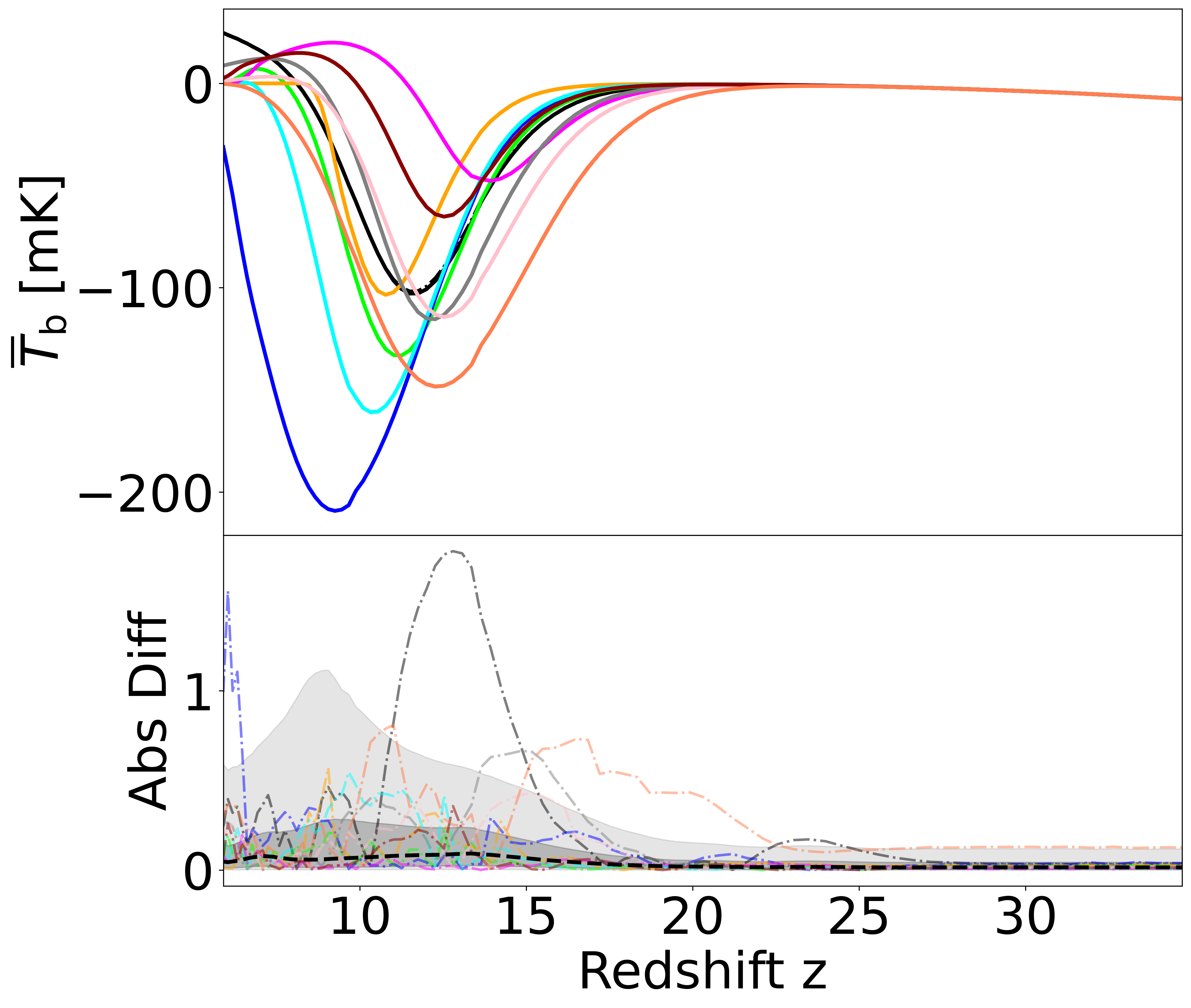} \\

    \includegraphics[width = 0.45\columnwidth]{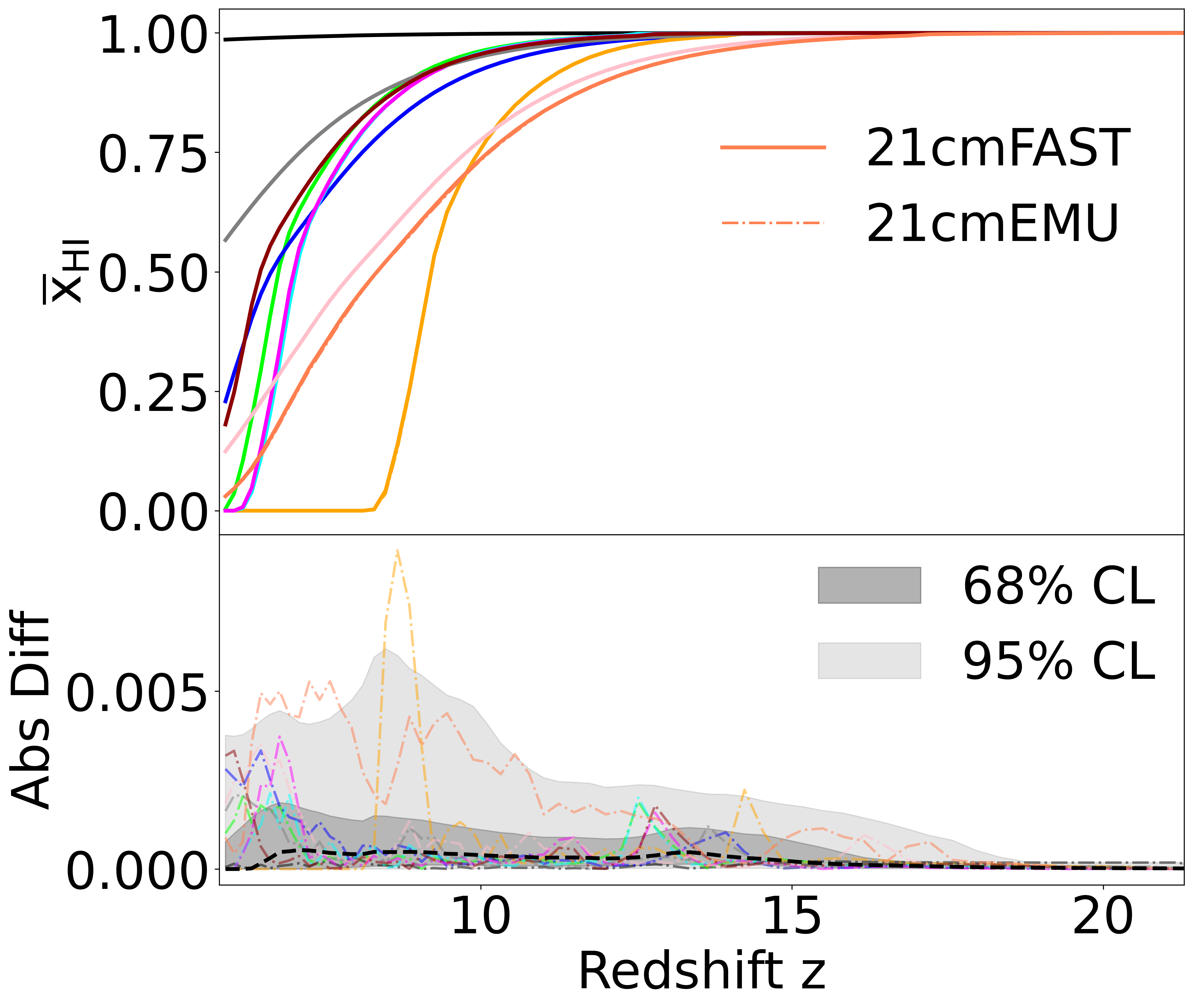}\hfill
    \includegraphics[width = 0.45\columnwidth]{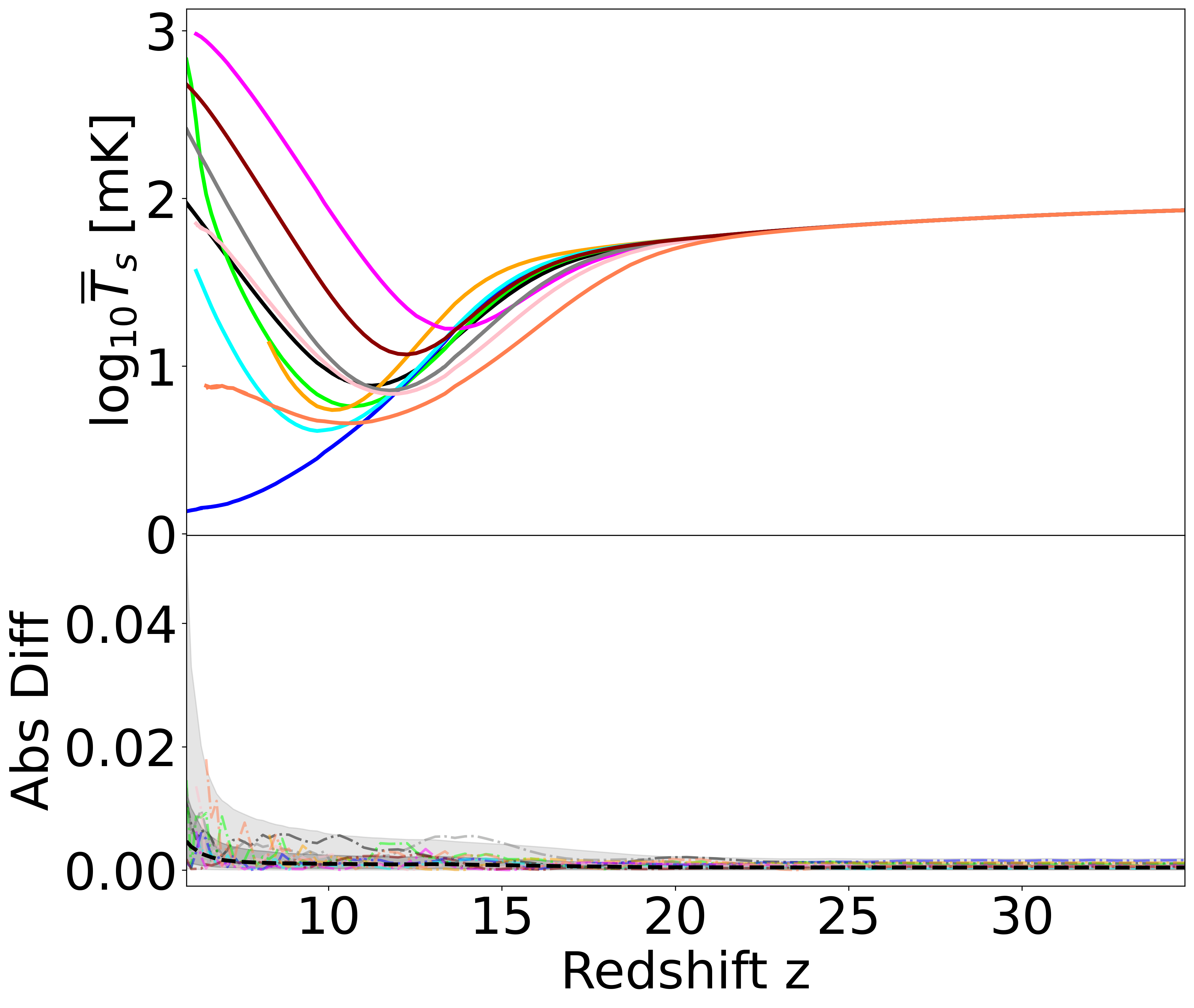} \\
    \includegraphics[width = 0.47\columnwidth]{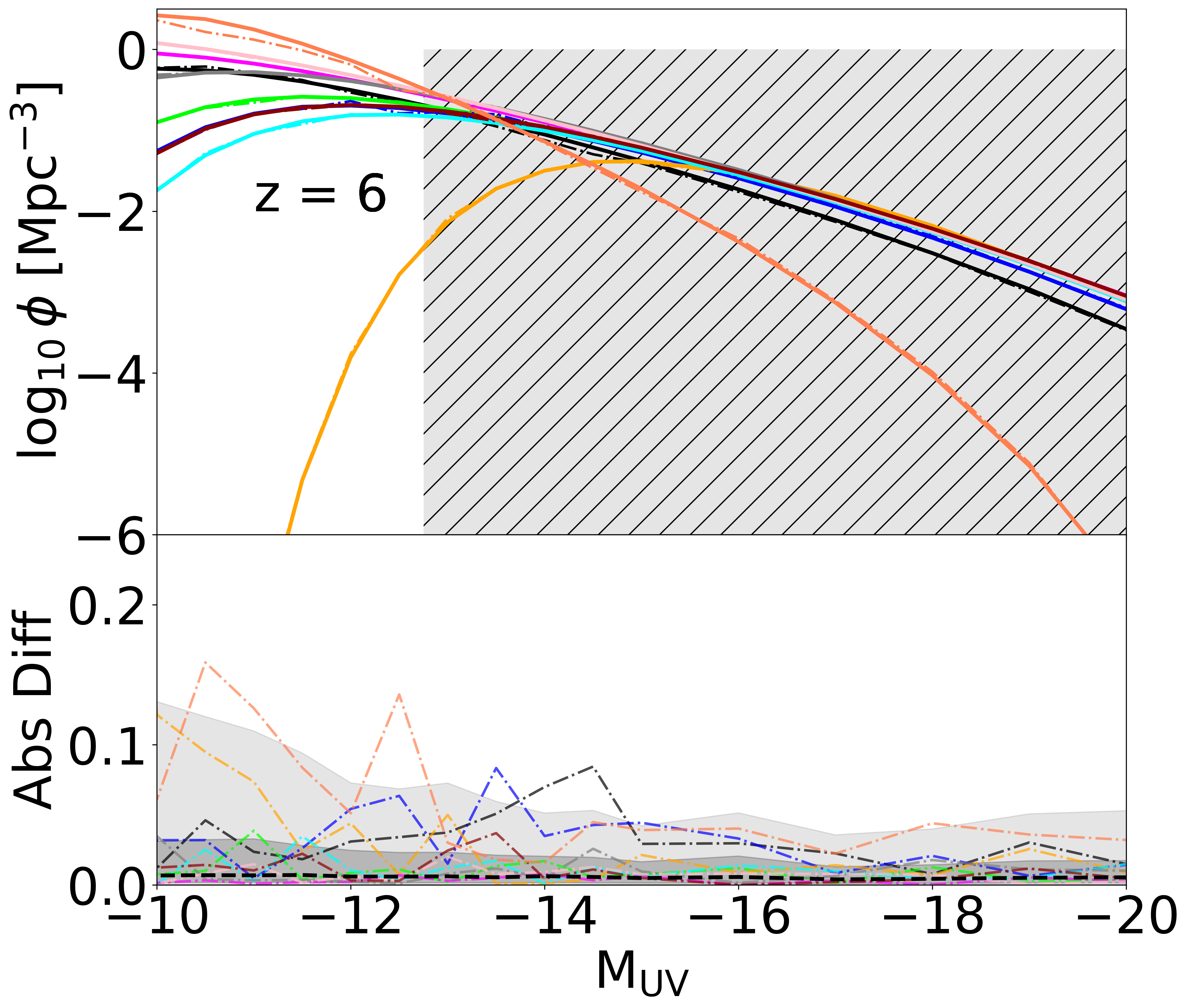} \hfill
    \includegraphics[width = 0.43\columnwidth]{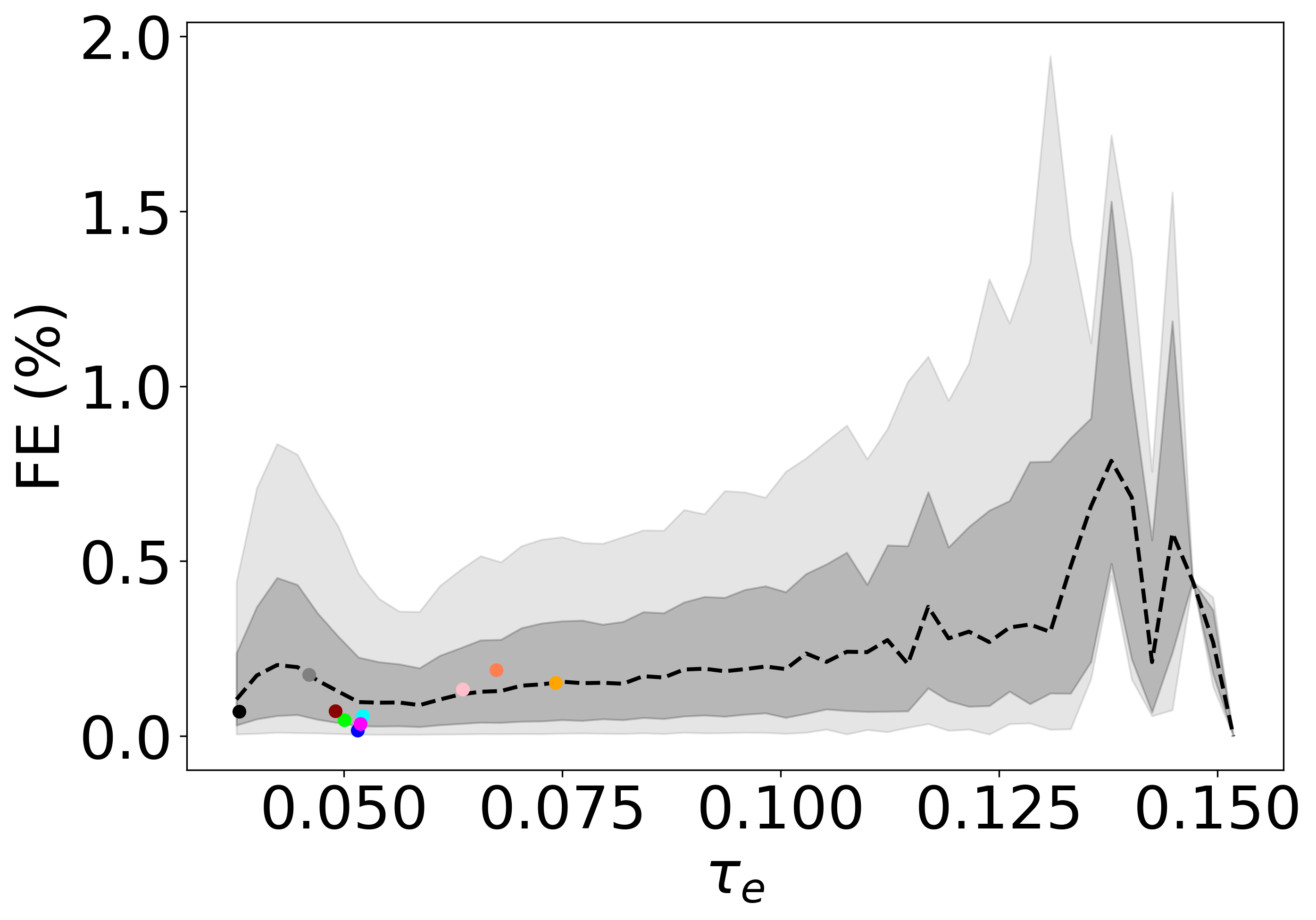}
    \end{subfigure}
    \caption{A subset of summary outputs from \cmemu\ for ten random samples from the test set.  Panels show: redshift evolution of the $k=0.1$ Mpc$^{-1}$ 21-cm PS amplitude, redshift evolution of the mean 21cm brightness temperature, redshift evolution of the mean spin temperature in the neutral IGM, the CMB optical depth, UV LFs at $z=6$, the EoR history ({\it clockwise from upper left}). Colors denote the astrophysical parameter sample with solid (dashed) lines corresponding to outputs from \cmfast\ (\cmemu).  In the bottom sub-panels, we show the absolute differences between the predicted and true quantities shown in the top sub-panels. Absolute differences of the ten random samples are shown with the corresponding colors, while the median absolute differences (FE in the case of $\tau_e$) computed over the entire test set are shown with dashed, black curves.   Dark (light) shaded regions enclose 68\% (95\%) CL.}
    \label{fig:emuvstrue}
\end{figure*}

We quantify the 21-cm PS prediction error 
in the top left panel of Figure \ref{fig:emuvstrue}.  In the top sub-panel, we plot the redshift evolution of the PS amplitude at $k=0.1$ Mpc$^{-1}$, with \cmemu\ predictions shown via dash-dotted lines and the corresponding \cmfast\ targets shown with solid lines.  We chose to plot $k=0.1$ Mpc$^{-1}$ because the strongest constraints by current interferometers are  around these scales; smaller scales are dominated by thermal noise and larger scales by foregrounds (e.g., \citealt{PSLOFAR,Trott20PSMWA,HERAJosh, HERAobs1}).
The ten models plotted here were chosen at random from the test set.  We again see that the differences between the emulator and "truth" are difficult to spot with the naked eye.

In the bottom sub-panel we 
show the absolute difference between each pair of curves in the top sub-panel, as well as the median absolute difference (dashed black line) and the 68\% $/$ 95\% confidence limits (CL; dark / light gray) computed over the entire test set.  We see that the median (68\%) \cmemu\ absolute error at $k\sim0.1$ Mpc$^{-1}$ is $\vline \ \log\left(\Delta^2_{21,\rm true} / {\rm mK}^2 \right)$ - $\log\left(\Delta^2_{21,\rm pred}/ {\rm mK}^2 \right)\vline \:$ $\leq$ 0.01 ($\sim$ 0.02).  This translates to a median (68\%) fractional error of 0.70\% (1.0\%) \footnote{Note that these errors are calculated on the emulator PS output which is in log space. Computing the corresponding error distributions in linear space, we obtain a median (68\%) FE of 1.53\% (1.94\%) at $k\sim0.1$ Mpc$^{-1}$, and 1.39\% (3.76\%) over the entire test set. Note that since we return to linear space, we do not need to apply a floor on the power spectrum in this FE calculation.}
at this wavemode and 0.55\% (2.4\%) when averaged over all wavemodes. This is far below observational uncertainties in the near-term, thus justifying the use of an emulator.  The error rises slightly at lower redshifts, owing to the broader distributions of possible PS, including very small values post reionization. 
In Appendix \ref{sec:worst}, we show the evolution of the 21-cm power spectrum fractional error as a function of the input 9D astrophysical parameters.

\subsection{The 21-cm global signal}\label{sec:tb}

The 21-cm global signal branch consists of seven fully-connected layers with 600 - 1000 nodes each. We quantify the performance of \cmemu\ on the global signal in the top right panel of Figure \ref{fig:emuvstrue}. We show the redshift evolution of the global signal ({\it top}) and absolute differences ({\it bottom}) for the same ten random samples from the test set.

As for the 21-cm power spectra, the difference between the \cmfast\ calculation and \cmemu\ prediction is difficult to see with the naked eye and is generally $\lesssim 1$ mK.  We see from the bottom sub-panel that the 95\% CL of the errors in the test set is also $\lesssim 1$ mK. This translates to a median (68\%) FE of 0.34\% (1.2\%). 

 We see from both the global signal and the PS that our training set spans a wide range of heating and ionization histories.  This is due to the fact that we include both accepted and rejected livepoints of the HERA22 inference in the training set, in order to have the largest dataset possible.  Extending beyond the ranges of the most likely models allows \cmemu\ to generalize beyond the HERA22 posterior distribution, accurately predicting even unlikely models that, e.g., have not reionized by $z=5$.

\subsection{The 21-cm spin temperature in the neutral IGM} \label{sec:ts}

The $\aveTs$ branch consists of five fully-connected layers with 400 nodes each. We quantify the network performance on the mean 21-cm spin temperature in the right panel of the middle row of Figure \ref{fig:emuvstrue}. In the top sub-panel, we show ten examples of the emulated spin temperature curve (dash-dotted line) and the corresponding true curves from the test set (solid line).
In the bottom sub-panel of the plot, we show the absolute error for each of the ten examples, the median for the entire test set with the black dashed line, and the 68\% $/$ 95\% CL regions in shaded in dark/pale grey as a function of redshift. We can see that the absolute difference is $\vline \: \log \big(\rm{T}_{\rm S, true} / \rm{K}\big) - \log \big(\rm{T}_{\rm S, pred} / \rm{K}\big)\vline \,$< 0.01 at 95\% CL over most of the redshift range. The FE of the log of the mean spin temperature over the entire test set is 0.032 \% and the 68\% CL is 0.13 \%.

We recall that the spin temperature is calculated by taking the global average over all cells in the simulation box that have $\xhi \geq 95\%$. When there are no cells satisfying this condition, the spin temperature becomes undefined.
We account for this by having the emulator predict the redshift at which the spin temperature becomes undefined.\footnote{In principle, one could use the EoR history emulator prediction to find the redshift at which the volume averaged neutral fraction drops below 0.05.  However, this is not identical to our definition for $\aveTs$, since our simulations account for partially neutral and self-shielded clumps inside the reionized cells.  Therefore we include a separate output for the redshift at which there are no cells with $\xhi \geq 95\%$. We note that \cmfast\ also includes partially-ionized cells, both by UV and X-rays.  Partial ionization by UV is assumed to correspond to unresolved HII regions surrounding nascent galaxies (see the discussion in \citealt{Zahn11}).}
The emulator correctly predicts the exact redshift bin below which $\aveTs$ becomes undefined for 95.1\% of the models in our test set, and is only one bin off ($\Delta z \sim 0.1$) for 4.89\% of the models.

\subsection{The global history of reionization}\label{sec:NF}

The EoR history branch, like the spin temperature branch, consists of five fully-connected hidden layers with 500 nodes each. In the left panel of the middle row in Figure \ref{fig:emuvstrue}, we show the EoR histories of our ten parameter samples ({\it top sub-panel}), and the corresponding prediction error ({\it bottom sub-panel}).  We see that the absolute differences are $\lesssim$0.005 for 95\% of the models in the test set.  The FE is 0.0075\% for the median and 0.095\% at the 68\% CL.

\subsection{The CMB Thomson scattering optical depth}\label{sec:taue}
The Thomson scattering optical depth branch consists of three layers of 30 nodes each as it outputs only one number. We show the FE of the $\taue$ prediction in the lower right panel of Figure \ref{fig:emuvstrue}. The ten parameter samples are denoted with different color dots.  Over the entire test set, we see a median fractional error of 0.1\% and a 0.25\% FE at 68\% CL. There is a notable increase in the prediction error as well as its bin-to-bin variance toward higher values of $\taue$. This is due to a small number of samples in this unlikely corner of parameter space: fewer than 1\% of the models in the test set have $\taue>0.11$.

\subsection{Galaxy UV luminosity functions}\label{sec:uvlfs}
The LFs branch consists of five layers of 400 nodes each. The network outputs the LFs at four redshifts ($\rm{z} = 6, 7, 8, 10$) and magnitude bins ranging from -20 to -10. 
In the lower left panel of Figure \ref{fig:emuvstrue}, we show the emulated and simulated LFs at $z=6$ (the performance at the other redshifts is comparable). The hatched region denotes the range spanned by LF observations used in the inference in the following section. 

We can see that the emulator is very accurate in the flat range spanned by the existing observations, while it is less accurate around the faint-end turnover. At all of the redshift bins, we have that the absolute difference $\vline \: \log \left(\phi_{\rm true}/\rm{Mpc}^{-3} \right) - \log \left(\phi_{\rm pred}/\rm{Mpc}^{-3}\right)\vline < 0.1$ over the majority of the magnitude range. 

We provide an alternative setting in \cmemu\ that allows the user to skip the emulation and directly calculate the CMB optical depth and UV LFs using \cmfast.  This improved accuracy however comes at the cost of a slower runtime:  $\sim$ 700 ms per call compared with $\lesssim$ 50 ms using emulation.

\subsection{Summary of \cmemu\ performance and context with other emulators}

In Table \ref{tab:network_summary}, we summarize the performance of \cmemu\ for each summary in the first row, using the fiducial training set of 1.3M samples.  In general, the median (68\%) emulation fractional error is at the level of $\lesssim$ 0.5\% (1\%).  The most accurate prediction is achieved with the EoR history, most likely due to the fact that it is a monotonic and smooth function, making it easier to learn. The least accurate summary is the power spectrum, which is understandable as it is two dimensional with the largest dynamic range.

It is difficult to directly compare the performance of \cmemu\ with other emulators of EoR/CD observables, due to their different astrophysical parametrizations and training set sizes.  Nevertheless, at face value \cmemu's accuracy is better than achievable with state-of-the-art emulators (e.g., \citealt{LOFARM20, GLOBALEMUB21, 21cmVAEBye22, emuYoshiuraEDGES}).  For example, comparing with the recent, bespoke 21-cm global signal emulator {\tt 21cmVAE} \citep{21cmVAEBye22}, we obtain a factor of 2.2 (1.5) lower median (95th percentile) RMS error (see their eq. 1).  Our median 21-cm PS FE is a factor of $\sim10$--100 lower than that of the bespoke PS emulators in \citealt{emuPSKern17, emuPSG20}, when compared over the same redshift/wavemode ranges.

This improvement in \cmemu\ over previous works could be attributed to several factors.  Firstly, we have a training set of unprecedented size: 1.3M samples.  This is orders of magnitude larger than used in previous works (generally ranging from thousands to tens of thousands).  We quantify how \cmemu's accuracy changes with the training set size in the following section.

Secondly, the improvement in power spectrum emulation could be attributed in part to our novel CNN architecture.  Previous 21-cm PS emulators used only fully-connected layers which are not as efficient in processing 2D images such as the PS.

Finally, the fact that \cmemu\ emulates many different observables allows the prediction of any one of these to be helped by the others.  Indeed, we verified explicitly that the 21-cm PS emulation is improved when the other summary outputs are included in the loss (i.e., when all branches are trained together).
In addition to improving performance, including multiple EoR/CD observables is extremely important in the current era where 21-cm observations are not strongly constraining.  As we show in Section \ref{sec:probes}, complementary galaxy and EoR observations are needed to obtain a likelihood-dominated (as opposed to prior-dominated) posterior (see also HERA22).

\subsection{Varying the size of the training set}
\label{sec:training_size}

Since \cmemu\ was trained on an uncharacteristically-large training set, it is useful to see how it performs with smaller training sets.  To do so, we remove some models at random, retrain \cmemu\ on the reduced training set, and test its performance on the same test set.


\begin{table}
\begin{tabular}{llll}
\hline
Training Size                               & \multicolumn{1}{c}{Summary} & \multicolumn{1}{c}{Median FE (\%)} & \multicolumn{1}{c}{68\% CL (\%)} \\ \hline
\multicolumn{1}{c|}{\multirow{4}{*}{\begin{tabular}[c]{@{}c@{}}1.3M\\ Full\end{tabular}}} & $\log\delsq$                          &  0.55                                  & 2.4                                 \\
\multicolumn{1}{c|}{}                       & $\aveTb$                &  0.34                                  & 1.2                                \\
\multicolumn{1}{c|}{}                       & $\log\aveTs$                &  0.032                                  &  0.13                                \\
\multicolumn{1}{c|}{}                       & $\xhi$               &  0.0073                                  & 0.10                                 \\ 
\multicolumn{1}{c|}{}                       & $\taue$               &  0.11                                  & 0.26                                 \\
\multicolumn{1}{c|}{}                       & $\log \phi$              &  0.50                                  & 2.1                                 \\ \hline
\multicolumn{1}{c|}{\multirow{4}{*}{\begin{tabular}[c]{@{}c@{}}640k\\ Random\end{tabular}}}  & $\log \delsq$                          & 0.71                                   &  3.0                                \\
\multicolumn{1}{c|}{}                       & $\aveTb$                &             0.43                       &   1.51                               \\
\multicolumn{1}{c|}{}                       & $\log \aveTs$                &             0.047                       &  0.17                             \\
\multicolumn{1}{c|}{}                       & $\xhi$               &             0.0086                       &  0.12\\ 
\multicolumn{1}{c|}{}                       & $\taue$               &  0.15                                  & 0.35                               \\
\multicolumn{1}{c|}{}                       & $\log \phi$              &  0.57                                  & 2.5                                 \\ \hline      

\multicolumn{1}{c|}{\multirow{4}{*}{\begin{tabular}[c]{@{}c@{}}13k\\ Random\end{tabular}}}  & $\log \delsq$                          &   3.2                                 & 13.0                                 \\ 
\multicolumn{1}{c|}{}                       & $\aveTb$                &        4.8                            &       16.6                           \\
\multicolumn{1}{c|}{}                       & $\log \aveTs$                &         0.40                          &       1.2                      \\
\multicolumn{1}{c|}{}                       & $\xhi$               &      0.035                              &    0.57 \\
\multicolumn{1}{c|}{}                       & $\taue$               &  0.45                             & 1.0                                     \\
\multicolumn{1}{c|}{}                       & $\log \phi$              &  2.5                                & 10.0                                
\end{tabular}
\caption{Performance of the \cmemu\ network when trained on the full database, half of the database and 1\% of the database.}
\label{tab:network_summary}
\end{table}

\begin{figure}
    \centering
    \includegraphics[width = \columnwidth]{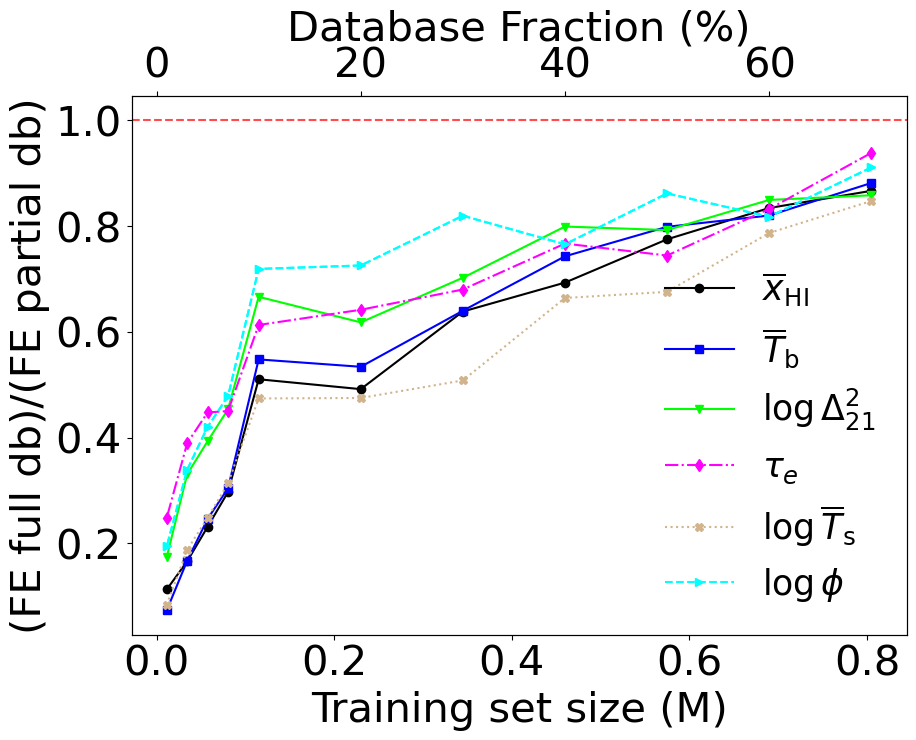}
    \caption{Median fractional error of each summary as a function of the training set size. 
 The FE is normalized so that unity corresponds to the fiducial, 1.3 M training set.}
    \label{fig:fe_vs_db}
\end{figure}

In Figure \ref{fig:fe_vs_db}, we plot the median FE in each summary as a function of the training set size.  We normalize the FE so that unity corresponds to the fiducial, 1.3 M training set.  We also explicitly list the performance using
half of the database (640k samples), and 1\% of the database (13k samples) in the middle and bottom rows of Table \ref{tab:network_summary}.

We see that there is a sharp increase in emulator accuracy with training set size, up to a size of $\sim$ 100k.  Doubling the size of the training set roughly doubles the emulator accuracy.  This relationship flattens beyond sizes of $\gsim$ 100k, such that a ten-fold increase in the training set from $\sim$100k $\rightarrow$ 1.3 M only improves the FE by a factor of $\sim$two.


\section{Application to inference} \label{sec:inference}

In this section, we apply \cmemu\ to inference problems. We use the {\tt 21cmMC} driver \citep{21cmMC15}, which now includes the option to use either \cmfast\ or \cmemu\ as the simulator. {\tt 21cmMC} incorporates three highly parallel samplers: EMCEE \citep{EMCEEFM13}, Multinest \citep{MultinestF09}, and Ultranest \citep{buchner16ultranest, buchner19ultranest, buchner21ultranest}; in this work we use the latter two as discussed further below.

First, we run the same inference as was previously run in HERA22 using \cmfast\ in order to see how emulator error affects the posterior. After this validation, we showcase the potential of \cmemu\ by performing several new inferences demonstrating: (i) how different observations are complimentary; (ii) the approximate impact of new, late-ending EoR constraints; (iii) the potential impact of upcoming H6C HERA observations. 
Each \cmemu\ inference took roughly a day on a GPU, compared with a few weeks on a cluster had we used \cmfast\ directly.


\subsection{Comparison with direct simulation}
\label{sec:21cmFASTinf}

We run the same inference as in HERA22 (10k livepoints) with \cmemu. Doing this allows us to directly compare the inference results between the two methods.

The likelihood in HERA22 incorporates four data sets:
\begin{enumerate}
\item {\it Thomson scattering CMB optical depth} - this term compares the Thomson scattering CMB optical depth from the proposed model with the one from the analysis of \citet{Planck18} by \citet{tauQ20}, whose posterior is characterized by median and 68\% credible interval (CI): $\tau_e = 0.0569^{+0.0081}_{-0.0086}$. The likelihood function is a two-sided Gaussian.
\item {\it The Lyman forest dark fraction} - this term compares the mean neutral fraction at $z = 5.9$ with the upper limit of $\avenf < 0.06 \pm 0.05$ at 68\% CI  obtained from QSO dark fraction \citep{DFMcGreer15}. The likelihood function is unity at $\avenf(z=5.9) < 0.06$, decreasing as a one-sided Gaussian for higher neutral fraction values.
\item {\it UV luminosity functions} -  this term compares the model with $z=6, 7, 8, 10$ UV luminosity functions observed with {\it Hubble} \citep{UVLFB15, UVLFB16, UVLFO18} in the magnitude range $M_{\rm UV} \in [-20, -10]$. This likelihood term is also a two-sided Gaussian.
\item {\it 21-cm power spectrum upper limits} - this term accounts for HERA H1C 94 night observations at $z=8$ and $z=10$, presented in \citet{HERAobs1}.  The likelihood is the upper limit likelihood discussed in HERA22.
\end{enumerate}
These individual likelihood terms are multiplied to obtain the total likelihood.
When using \cmemu\ for inference, we add the median emulator error in quadrature to the measurement uncertainties for each corresponding likelihood term.

\begin{figure*}
    \centering
    \includegraphics[width =  \textwidth]{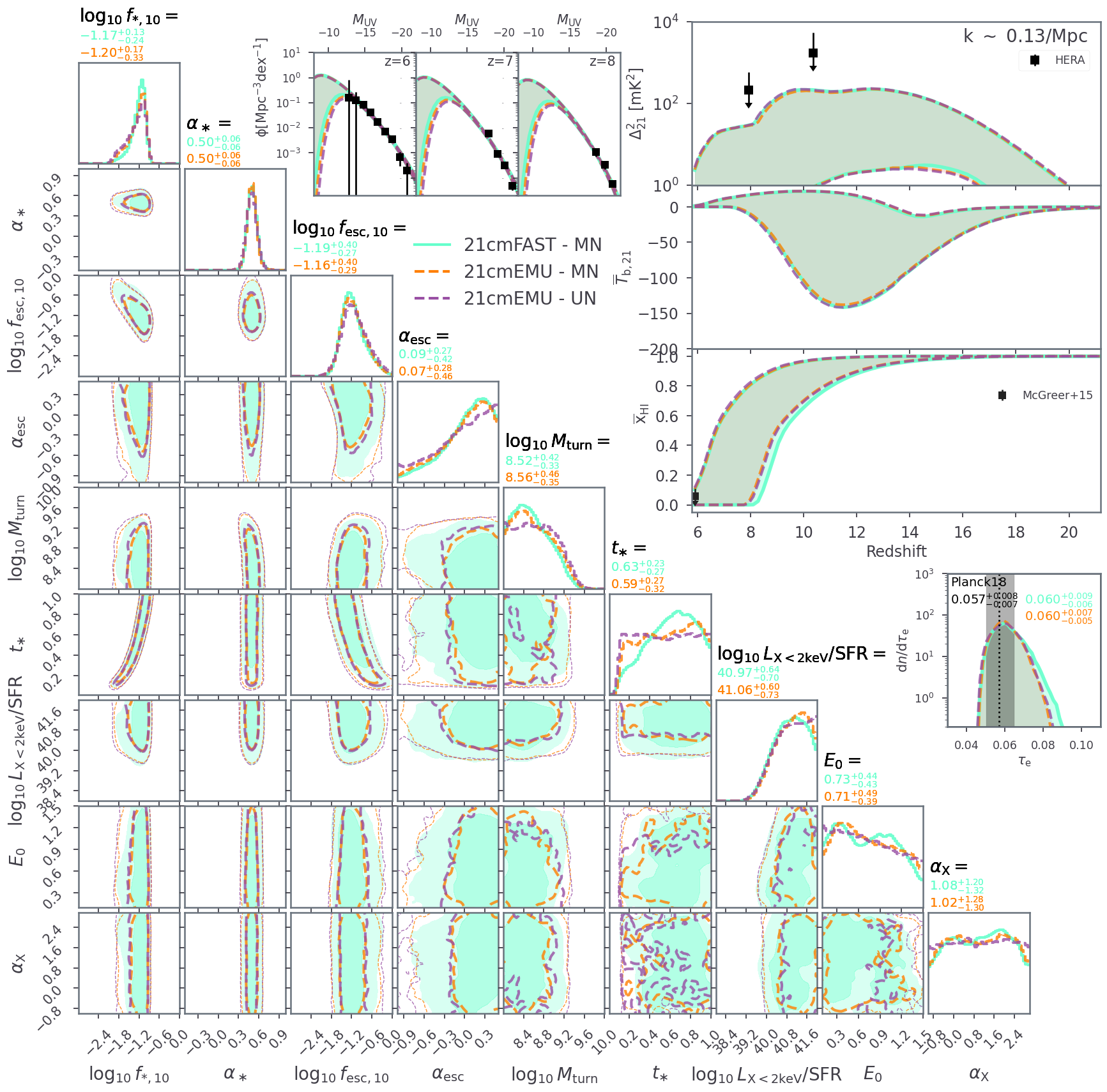}
    \caption{Comparison of posteriors obtained using \cmfast\ and \cmemu\ after performing an inference with the same HERA22 likelihood. The darker/dashed regions represent 68\% CIs, while pale/thin dashed regions represent 95\% CIs. The orange and purple posterior distributions are obtained using the \texttt{MultiNest} sampler (10k livepoints, $\sim 60$k posterior samples), while the cyan posterior distribution is obtained using the \texttt{UltraNest} sampler (5k livepoints, $\sim$ 70k posterior samples). The median value and the 68\% CIs of the 1D marginal PDFs are written above each column of the corner plot.
    In the panels on the top right, all highlighted regions correspond to 95\% CIs. In the top middle panel, we plot the luminosity functions for redshifts 6,7, and 8. For the luminosity function likelihood, we use the data shown in black squares \citep{UVLFB15, UVLFB16,UVLFO18}. In the top right, we show a panel with three summary statistics, namely the redshift evolution of the 21-cm power spectrum at $k = 0.13$ cMpc$^{-1}$, the 21-cm global signal and mean neutral fraction, from top to bottom. The black squares in the power spectrum plot correspond to the two deepest limits for each HERA redshift band ($k = 0.13$ cMpc$^{-1}$ at z $\sim8$ and $k = 0.17$ cMpc$^{-1}$ at z $\sim10$). In the bottom plot, the black square denotes the upper limit on the average neutral hydrogen fraction obtained from the QSO dark fraction \citep{DFMcGreer15}. In the bottom right, we show the PDFs of the Thomson optical depth together with the {\it Planck} result used in the likelihood. The astrophysical parameter ranges shown in the corner plot correspond to the extent of the flat priors assumed for the inferences.}
    \label{fig:corner_compare}
\end{figure*}

\begin{figure*}
    \centering
    \includegraphics[width =  \textwidth]{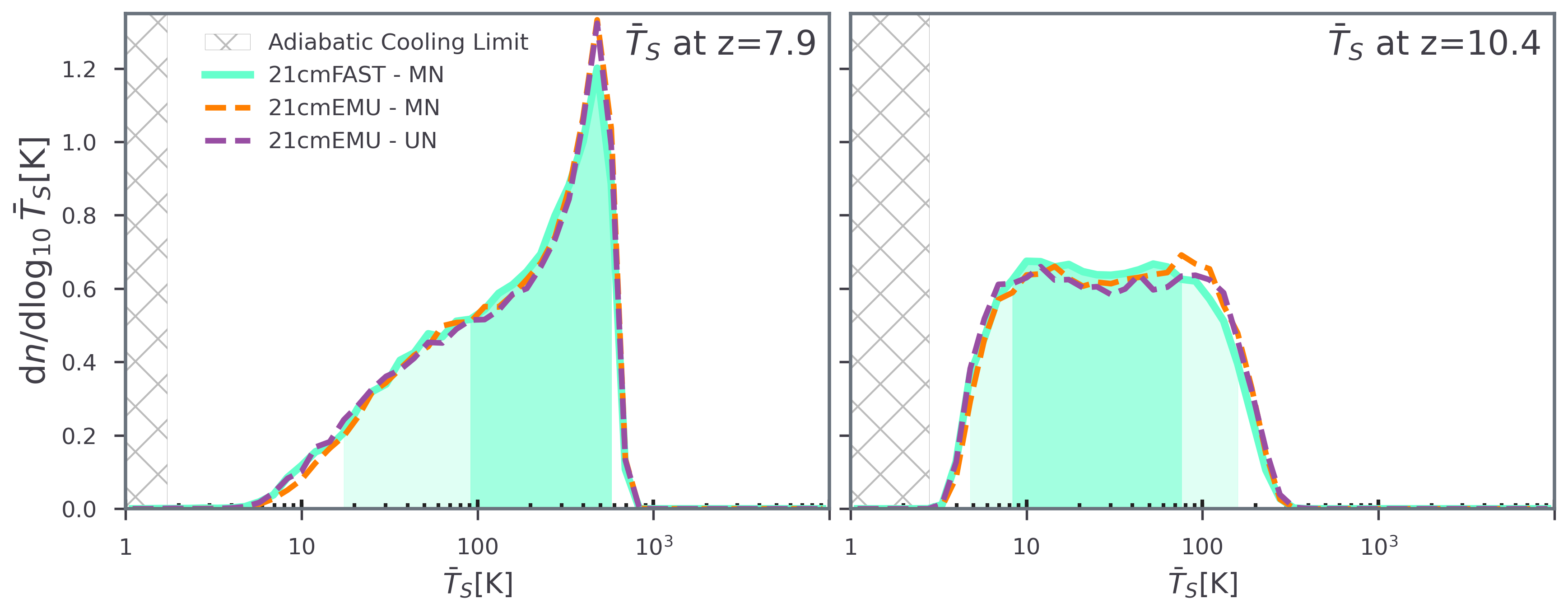}
    \caption{Comparison of the mean spin temperature distribution from \cmfast\ and \cmemu\ for each of the two HERA bands after performing an inference with the exact same likelihood. The credible intervals have been calculated using the highest posterior density method. The dark (light) cyan shaded region shows the 68\% (95\%) CI. The solid cyan line shows the distribution for \cmfast\ with 10k livepoints using \texttt{MultiNest}. The dashed orange line shows the same but for \cmemu . The dashed purple line shows the distribution for \cmemu\ but using the \texttt{UltraNest} sampler with 5k livepoints.}
    \label{fig:Ts_compare}
\end{figure*}

In Figures \ref{fig:corner_compare} and \ref{fig:Ts_compare}, we compare posteriors obtained using \cmfast\ ({\it cyan}) to that using \cmemu\ ({\it orange}).  Both were run using the \texttt{MultiNest} sampler with the same number of livepoints (10k, yielding $\sim 60$k posterior samples).  In the lower left of Fig. \ref{fig:corner_compare} we plot the 1D and 2D marginal PDFs for our astrophysical parameters, while in the top right we plot 95\% CI of some of the summary observations (see caption for details).  In Fig. \ref{fig:Ts_compare} we plot the corresponding spin temperature PDFs in the two HERA bands, which was one of the main results of the HERA22 paper.
We note that the \cmfast\ and \cmemu\ posteriors are nearly identical, testifying that the emulation error is fairly negligible when performing inference using current data sets.   The only notable difference is in the $t_\ast$ PDF, which is slightly broader when \cmemu\ is used as a simulator compared with \cmfast.  We find no notable trends of the emulator error with this parameter, concluding the small difference could be due to stochasticity in sampling and/or a higher dimensional covariance of the emulator error.

In Figure \ref{fig:corner_compare} we also include a run using \cmemu\ and the same HERA22 likelihood, but with the \texttt{UltraNest} sampler ({\it purple curves}; 5k livepoints, yielding $\sim 70$k posterior samples).  The resulting posterior is consistent with the previous two.  Interestingly, the choice of sampler (purple vs orange) results in a larger difference than the choice of simulator (orange vs cyan) for some marginal PDFs.  In particular, the \texttt{UltraNest} posterior is more accurate towards the edges of the prior range, resulting in flatter posteriors at the edges: this behavior is also recovered using the \texttt{EMCEE} sampler as shown in \citet{Hovav23}.  Moreover, \texttt{UltraNest}'s vectorization makes it $\sim10$x faster when using an efficient simulator like \cmemu.  Therefore, in subsequent sections we only show posteriors generated with \texttt{UltraNest}.

We remind the reader that the emulator was trained on the HERA22 nested sampling output.  This inference took $\sim$400k core hours. Once trained however, the emulator performs amortized posterior estimation in only 225 core hours using \texttt{Multinest} or in 30 core hours using \texttt{Ultranest}.

\subsection{Impact of different observations on the posterior} \label{sec:probes}

\begin{figure*}
    \centering
    \includegraphics[width =  \textwidth]{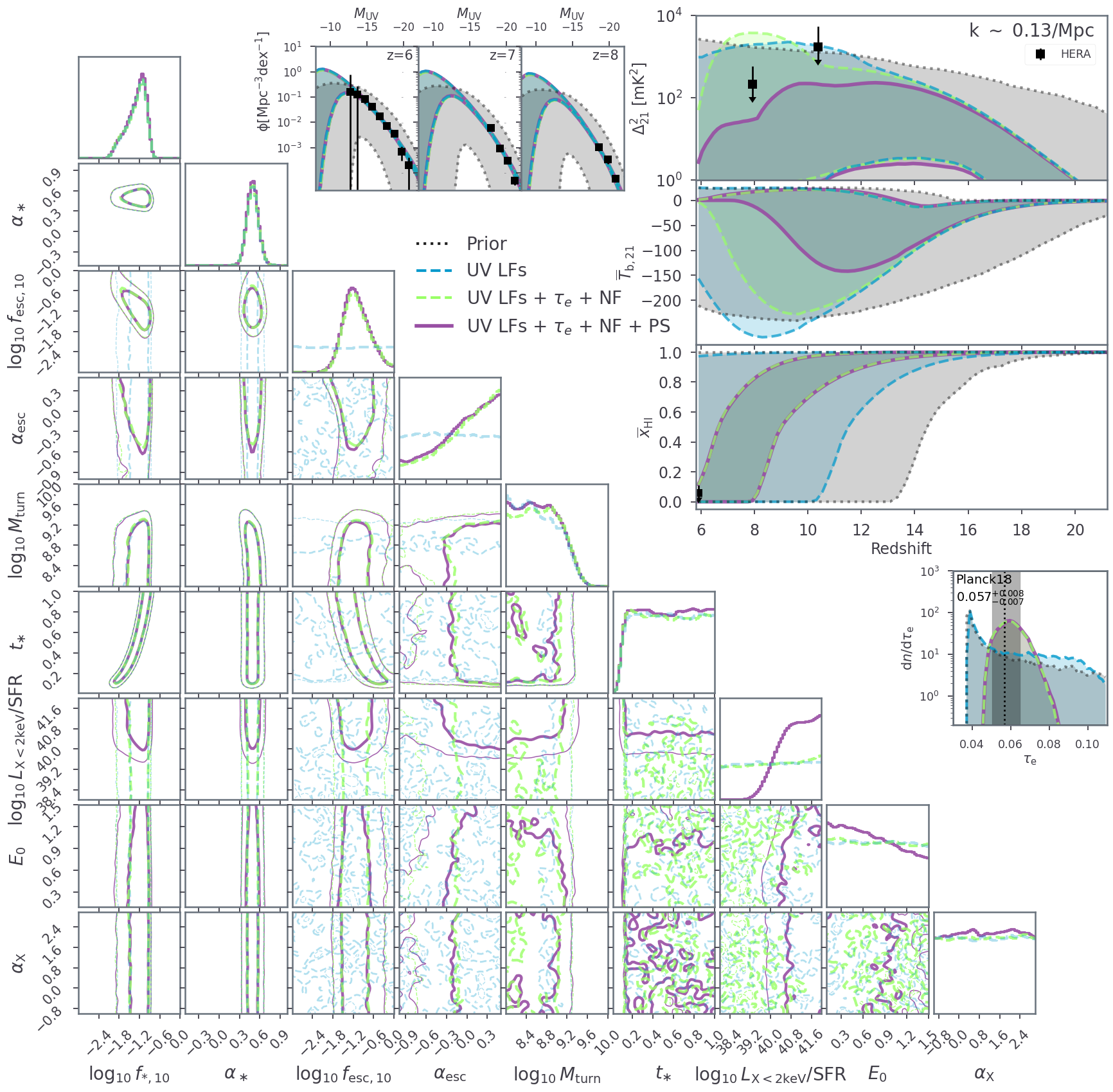}
    \caption{Contribution of various likelihood terms to the total posterior. The corner plot on the left shows the 95\% CI of three inferences, all run with \cmemu\ and \texttt{UltraNest}. The full posterior with all four probes is plotted in purple (exactly the same as the purple in Figure \ref{fig:corner_compare}). In green, we show the posterior without the HERA power spectrum upper limits term. In blue, we additionally remove the neutral fraction and Thomson optical depth terms, leaving only the UV luminosity functions terms. On the top right half of the plot, we show the 95\% CI of the same three posteriors but in the space of summary statistics: first the UV LFs, and then a panel with the 21-cm power spectrum, 21-cm global signal, and EoR history, top to bottom, and finally a panel with the Thomson optical depth. In grey, we plot the summary statistic 95\% CI assuming a flat distribution across all nine astrophysical parameters which is what was used for the prior for the \cmfast\ inference.}
    \label{fig:corner_probes}
\end{figure*}

Having tested the emulator in the previous section, we now use it to perform multiple inferences 
that would be too costly with direct simulation.
We begin by quantifying  how the individual terms from the HERA22 likelihood discussed in the previous section affect the posterior. We do this by removing the terms one by one, and comparing the resulting posteriors in Figure \ref{fig:corner_probes}.

In orange we show the full HERA22 posterior from the previous section, including all likelihood terms.  In green, we remove the HERA power spectrum upper limit constraint. We see that the only consequence is that the $L_X/{\rm SFR}$ parameter becomes unconstrained. In the panel on the right, we can also see the 95\% CI of the power spectrum and 21-cm global signal becoming wider around z $\sim 6-10$.  As discussed in HERA22, the 21-cm power spectrum limits is the only measurement sensitive to the IGM temperature during the cosmic dawn.

Next, if we remove constraints on the EoR history (here corresponding to the dark fraction and $\tau_e$ likelihood terms), using only the UV LFs in the likelihood, we obtain the posterior shown in blue.  We see that EoR history measurements allow us to set (lose) constraints on the ionizing escape fraction (here parametrized via $f_{\rm{esc},10}$ and $\alpha_{\rm{esc}}$), which disappear completely when their corresponding terms are not included in the likelihood. Including only the UV LFs does disfavor very early reionization, $z>11$, because the redshift evolution of the SFR density implied by UV LF observations is too steep to allow arbitrarily early EoR, even with escape fractions close to unity.


Finally we show the prior distribution in the space of UF LFs, 21-cm PS, 21-cm global signal, and EoR history in gray.  We see that all of the posteriors in these spaces are significantly broader than the priors, and are thus likelihood dominated (i.e., are not sensitive to the prior choices).  Moreover, each likelihood term adds complimentary information, highlighting the importance of combining observational datasets when interpreting the high-redshift Universe.

\subsection{Impact of late reionization}
\label{sec:lya}

Recent observations of the large-scale opacity fluctuations in the Lyman alpha forest (e.g., \citealt{quasarsBecker15, LyaFBosman18,Bosman22}) imply a late end to reionization $z<5.6$ \citep{LOFAR21,EoRLyaQin21}.  In this section, we explore how such new EoR history constraints would impact the previously-shown posterior.  Unfortunately, the current version of \cmemu\ does not emulate the Lyman alpha forest, and so we cannot compute a likelihood directly in the observed space of Ly$\alpha$ opacity fluctuations.  Instead we take a more approximate approach, computing the likelihood in the space of EoR histories, i.e., $\avenf(z)$. To construct a likelihood in this space, we use the EoR history posterior from \citet{EoRLyaQin21}, who did in fact compute a likelihood from forward-modeled Ly$\alpha$ opacities in addition to the dark fraction and $\taue$ observations.  Specifically, we compute a new {\it Late EoR} posterior by replacing the dark fraction and $\taue$ likelihood terms with a Gaussian likelihood evaluated at three redshifts, $\avenf(\rm{z} = 6) = 0.25 \pm 0.07$, $\avenf(\rm{z} = 7) = 0.58 \pm 0.1$, and $\avenf(\rm{z} = 8) = 0.79 \pm 0.09$, ignoring any covariance between redshifts.  Although this is obviously an approximation to computing the likelihood directly in the space of the observations, it suffices to qualitatively show the impact of new EoR history constraints.

\begin{figure*}
    \centering
    \includegraphics[width =  \textwidth]{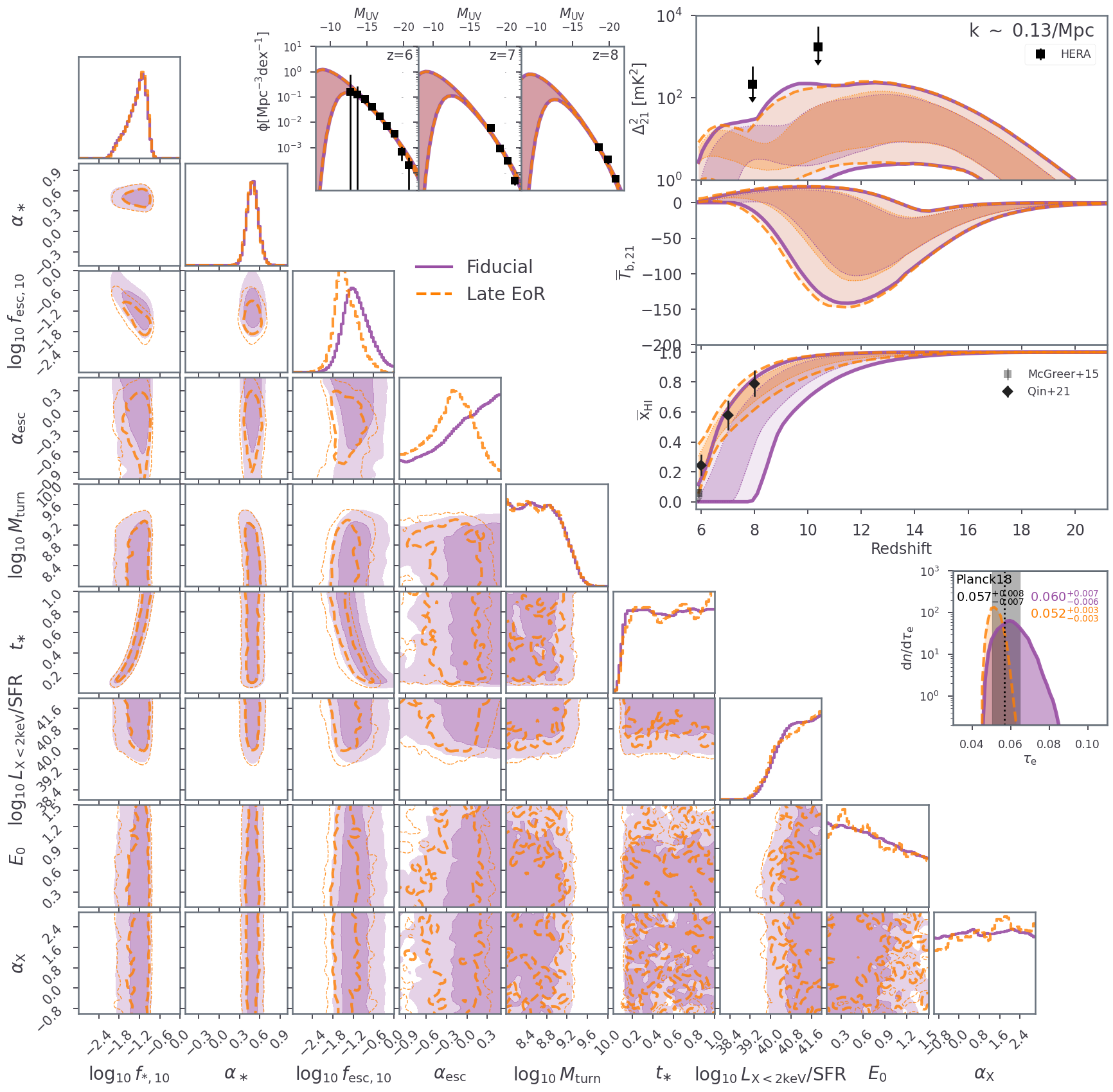}
    \caption{Same as Fig. \ref{fig:corner_compare}, but comparing the fiducial posterior ({\it purple}) with one obtained by replacing the QSO dark fraction and $\tau_e$ likelihood terms with a "Late EoR" likelihood denoted by the three points with error bars in the middle right panel ({\it orange}.  The "Late EoR" likelihood is based on the inference results in \citealt{EoRLyaQin21}, which included recent measurements of opacity fluctuations in the Lyman alpha forest.  In the top right sub-panels, we show both the 68\% (darker) and 95\% (paler) C.I.}
    \label{fig:corner_lya}
\end{figure*}

In Figure \ref{fig:corner_lya}, we show the previous ({\it Fiducial}) posterior in purple (71k samples) together with the new ({\it Late EoR}) posterior in orange (18k samples).  Understandably, the corresponding recovered EoR history in orange is narrowly centered around the three points at $z$=6, 7, 8 used to define the likelihood.  As a consequence, the posterior of the Thomson optical depth also becomes more narrow, shifting toward lower values while still being within the range allowed by Planck observations. The resulting PDF of $f_{\rm esc, 10}$ for {\it Late EoR} is also narrower, and shifted towards smaller values.  Even the power law scaling of the escape fraction with halo mass, $\alpha_{\rm esc}$, is constrained to within $\pm$ 0.3 (68\% C.I.) for {\it Late EoR}, whereas the {\it Fiducial} posterior only sets a lower limit for this parameter.  The remaining parameters are unaffected by the change to the Late EoR likelihood. 

We also see that the recovered 21-cm large-scale PS for {\it Late EoR} is narrower at $z<8$.  The large-scale 21-cm PS during the EoR peaks around its midpoint (e.g., \citealt{Lidz07, Pritchard07}), which occurs at $z\sim$7--8.  The HERA22 upper limits disfavor higher values of the 21-cm PS at $z\sim8$, but the tail towards small PS values seen in the {\it Fiducial} posterior (corresponding to small $\avenf$), shrinks when moving to the {\it Late EoR} posterior.


\subsection{Forecasts for HERA Phase II sixth-season observations}

We now forecast parameter constraints that could be achievable with the sixth season of HERA observations, taken in 2022-2023 (Berkhout \textit{et al.}, \textit{in prep.}).
This season of observing used Phase II of the HERA instrument, spanning 50-230 MHz (omitting the FM band, 90-110\,MHz), expanding coverage to Cosmic Dawn and late reionization with respect to Phase I (which was used for HERA22). 
While analysis of this season's data is ongoing, its broad characteristics are known \citep{HERAMemo122}: approximately 1300 hours of unflagged data over $\sim 150$ nights, with an average of $\sim 148$ un-flagged antennas per night. Although further flagging will certainly occur during processing, this dataset will be HERA's most sensitive data release to date, by a significant factor. 

We create a mock observation corresponding to this upcoming dataset.  For the "true" cosmic signal, we use the Evolution of Structure (EOS) 2021 release \citep{EOS2021}. EOS2021 is a large simulation (1.5 cGpc per side with $1000^3$ cells) made with \cmfast, with the goal of being our current "best guess" for the true cosmic signal. Although it used the same parametrization for galaxy scaling relations as is used here (see Section \ref{sec:astro_params}), the physical model of EOS2021 has a few notable differences.  Instead of leaving $M_{\rm turn}$ as a free parameter, EOS2021 explicitly calculated a local $M_{\rm turn}({\bf x}, z)$ based on feedback from the local ionizing and photo-disassociating backgrounds, as well as the relative velocities of baryons and dark matter.  Furthermore, EOS2021 explicitly accounted for putative PopIII star formation in the first, H$_2$-cooled galaxies (MCGs, e.g., \citealt{Tegmark97, Abel02,Bromm04, Haiman06}), which dominated the background radiation fields at $z>16$, for their fiducial parameter choices.
As a result of the models being different,  \cmemu\ could result in a biased recovery of the EOS2021 signal; we quantify this below.

We use \texttt{21cmSense}\footnote{https://github.com/rasg-affiliates/21cmSense} \citep{Pober13,Pober14} to obtain thermal and sample variance estimates of the HERA 6th season data, and describe our methodology and assumptions in App. \ref{app:sense}. We consider our sensitivity estimate to be realistic, with a few important caveats, for example the potential over-estimation of sensitivity when treating `similar' baselines as identical \citep{Zhang2018}.
The largest unpredictable caveat is of course the presence of instrumental systematics, for which we describe our approach in more detail below.

%


Radio telescopes, including HERA, impose their own signature on observations -- dependent on the primary beam attenuation, antenna layout, channelization and other instrumental characteristics. The effect of this instrumental signature on the observed power spectrum is such that neighbouring Fourier modes are linearly 'mixed' via a `window function' matrix (e.g., \citealt{Liu11,Gorce23}). 
We calculated this window function  using the \texttt{hera\_pspec}\footnote{\url{https://github.com/HERA-Team/hera_pspec}} package. 
We did not use the exact HERA beam as in \citealt{Gorce23}. Instead, we used the Gaussian beam approximation which we deemed sufficient for this forecast (see Figure 7 in \citealt{Gorce23} for a comparison). Once we obtain the HERA window function, we matrix multiply it with the emulated model to properly compare with the forecast.

\begin{figure*}
    \centering
    \includegraphics[width =  \textwidth]{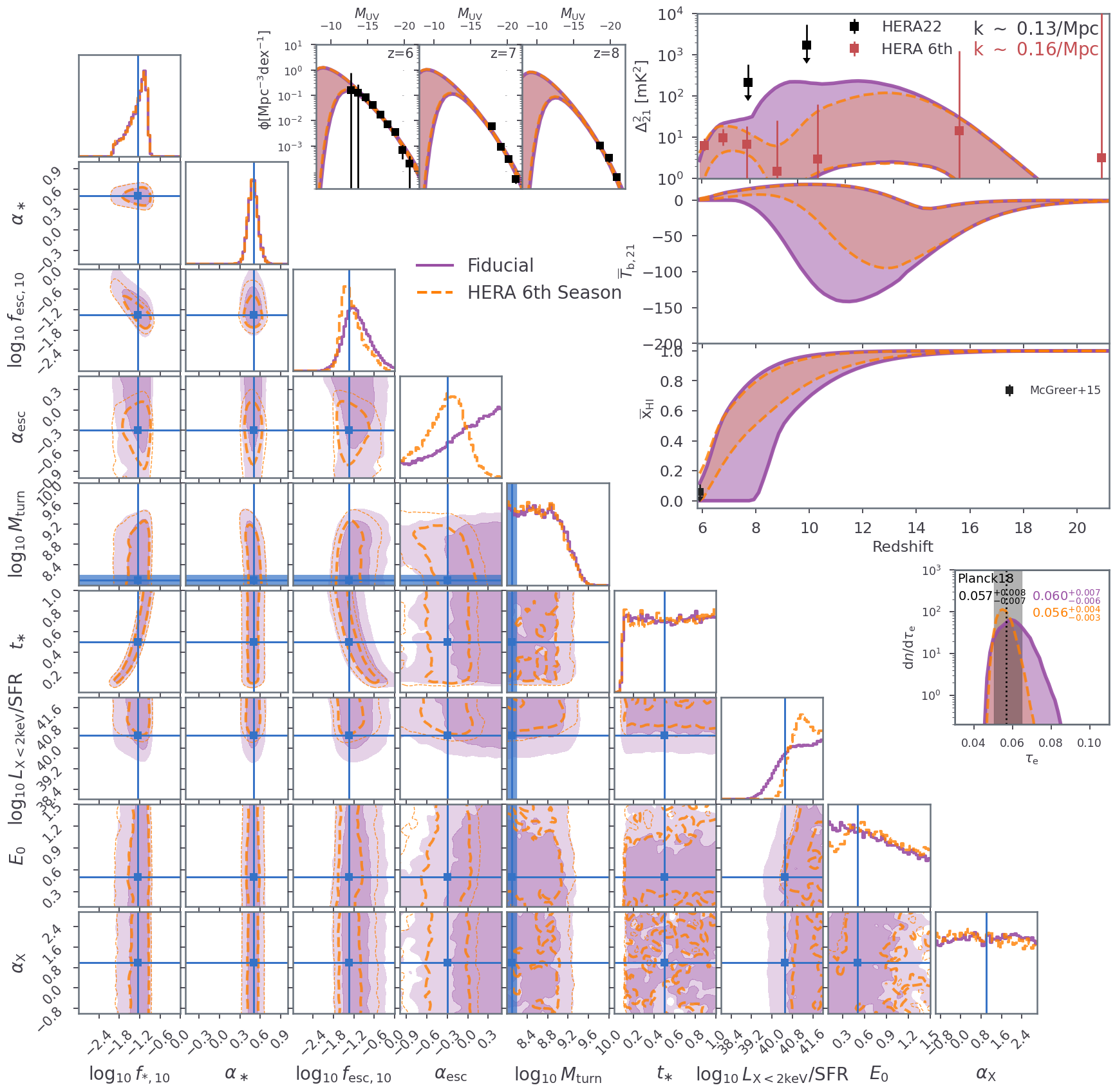}
    \caption{
    Forecasted constraints from mock HERA Phase II season six observations (see text for details) are shown in orange.  The mock PS amplitudes at $k\sim0.16$ Mpc$^{-1}$ are shown as orange points with error bars in the top right panel, together with current upper limits from HERA22.  
 The parameters of the cosmological simulation used for the mock observation, EOS2021, are denoted with blue lines in the corner plot.  We caution that the theoretical model used in EOS2021 and that used in \cmemu\ are somewhat different, as discussed in the text.  As $M_{\rm turn}$ in EOS2021 evolves with redshift (see Fig. 5 in \citealt{EOS2021}), here we demarcate its range during the EoR (i.e., $6<z<8$ where the mock observations imply a detection).
For more details about the panels, see the legend in Fig. \ref{fig:corner_compare}.}
    \label{fig:corner_h6c}
\end{figure*}

We perform inference using the EOS2021 cosmological signal with the sensitivity estimates from \texttt{21cmSense} as the mock observation (see Fig. \ref{fig:H6C_sens} in Appendix \ref{app:sense}). 
This inference takes about 30 GPU h to run to convergence with \texttt{UltraNest}.  In Fig. \ref{fig:corner_h6c} we show the resulting posterior ({\it HERA 6th season} in orange) together with the previous result ({\it Fiducial} in purple).
In the top right panel we show the mock PS at $k\sim0.16$Mpc$^{-1}$ as orange points with associated error bars.  We see that based purely on the available S/N, the HERA sixth season data have the potential to detect the cosmic PS during the EoR ($6<z<8$).  The 95\% CI of the inferred PS ({\it orange}) go tightly around the data points.  This unbiased recovery is reassuring, given the above-mentioned differences in the theoretical models used for the mock and forward-modeled data.  Indeed, most of the "true" astrophysical parameters from EOS2021 (denoted with blue lines in the corner plot) are consistent with the recovered orange posteriors.
Parameters governing X-ray heating, $L_{\rm X<2keV}/{\rm SFR}$ and $E_0$, are recovered with the lowest accuracy, with the true values residing outside of the 68\% CI of their 2D PDF.  This is understandable, because the \cmemu\ forward models do not include the additional radiation from $H_2$-cooling galaxies, which dominate the X-ray heating at $z>16$.
%


Comparing to current constraints ({\it Fiducial} posterior in purple), we see that that HERA 6th season data have the potential to drastically improve our knowledge of the EoR.  The {\it HERA 6th season} EoR history $\avenf(z)$ is constrained to within $\pm$0.1 (95\% C.I.): a factor of $\gtrsim 3$ improvement over current limits.  As a result, we can place much stronger constraints on the characteristic ionizing escape fraction, $f_{\rm esc, 10}$, and its dependence on galaxy mass, $\alpha_{\rm esc}$, which are almost completely unknown currently.

It is important to note that these two posteriors use a different form for the likelihood.
For the {\it HERA 6th season} forecast, we  assume that there are \textit{no residual systematics} after processing of the HERA data.
This is in contrast to the previous likelihoods, which assume that each $k$-mode has a positive systematic whose prior amplitude is uniform and unbounded \citep[cf.][]{HERAobs1}. In practice, assuming no residual systematics results in a two-sided Gaussian likelihood, corresponding to a `detection', whereas the traditional likelihood has been a one-sided error-function corresponding to an `upper-limit'. 
We make this choice as it is not straightforward to 
sample from the unbounded uniform prior for systematics when creating the mock data for the forecast.
The resulting tighter parameter posteriors for the new data are therefore the result of an admixture of the new more sensitive data \textit{and} the (effectively) more constrained priors on systematics.



\section{Conclusion} \label{sec:conclusion}

Here we introduced \cmemu: a publicly-available emulator of several summary observables from \cmfast.
We trained the emulator on 1.3M pseudo-posterior samples from the inference in HERA22. 
The input consists of a nine parameter model characterizing the UV and X-ray outputs of high redshift galaxies.  The output consists of: (i) the 21-cm power spectrum as a function of redshift and wavemode; (ii) the IGM mean neutral fraction as a function of redshift; (iii) the UV luminosity function at four redshifts 6, 7, 8, and 10; (iv) the Thompson scattering optical depth to the CMB; (v) the mean spin temperature as a function of redshift; and (vi) the 21-cm global signal as a function of redshift. The emulator predicts all of these quantities with under $\sim 2.4 \%$ error at 68\% CL, and a computational cost that is lower by a factor of $\sim$10000 compared to \cmfast.

We varied the size of the training set, finding only a modest decrease in performance (a factor of $\sim$2 decrease in the FE) as the number of samples was reduced from 1.3M to $\sim$100k.  Below $\sim$100k samples, we saw a sharp drop in performance, with the fractional error increasing roughly as the inverse of the size of the training set.

We validated the emulator's performance in inference by comparing the posteriors obtained with \cmemu\ vs \cmfast\, using the same likelihood (taken from HERA22).  We found a very modest difference between these two posteriors, further illustrating that the emulator error is negligible when performing inference using current data.

Next, we profited from the speed of our trained emulator to perform multiple inferences that would otherwise be very costly using direct simulation.  First, we studied the constraining power of each term in our fiducial likelihood.  We found that current observations are very complementary, with UV LFs constraining the stellar-to-halo mass relations, EoR history probes constraining the ionizing escape fraction, and the addition of 21-cm PS upper limits constraining the X-ray luminosity to SFR relation.

We also explored the impact of new EoR history constraints, driven by opacity fluctuations in the Lyman$\alpha$ forest.  These recent observations imply much tighter constraints on the EoR history, finishing at $z<5.6$ (e.g., \citealt{LyaQ21, LOFAR21}).  The inclusion of these new limits tightened the recovered constraints on the ionizing escape fraction and its scaling with halo mass.  The impact on other parameters was modest.

Finally, we presented forecasts of parameter constraints achievable with ongoing 6th season phase II observations with the HERA telescope.  Optimistically, we could expect a detection of the 21-cm PS at $z\sim$6--7.  This would result in a large improvement in the recovered timing of the EoR, allowing us to infer $\avenf(z)$ to within $\pm$ 0.1 (95\% C.I.): a factor of $\gtrsim 3$ improvement over current limits.  As a result, we could place stronger constraints on the characteristic ionizing escape fraction and its dependence on galaxy mass, which are almost completely unknown currently.  We cautioned however that this forecast is optimistic, mainly because it assumed there are no residual systematics in the processed data (see Appendix \ref{app:sense} for more details).

\cmemu\ was trained on a database of summary observables where only one seed i.e., random set of initial conditions is available per combination of astrophysical parameters. In the future, we hope train the emulator on a database that samples many different seeds in order to emulate a full likelihood function rather than only approximate the mean as we do right now. This is important since \citet{Prelogovic23} showed that using a single random seed when forward modeling can bias the inference results. 

We make \cmemu\ publicly-available at \url{https://github.com/21cmfast/21cmEMU}, and include it as an alternative simulator in the public \cmmc\footnote{\url{https://github.com/21cmfast/21CMMC}}\ sampler.  We will periodically release updated versions, trained on the latest galaxy models and expanding the choice of summary outputs.

\section*{Acknowledgements}
We thank A. Liu and J. Dillon for useful comments on a draft version of this paper.
We also thank A. Gorce for helpful discussion concerning the HERA window function calculation.
We gratefully acknowledge computational resources of the Center for High Performance Computing (CHPC) at SNS.
A.M and R.T. acknowledge support from the Italian Ministry of Universities and Research (MUR) through the PRO3 project "Data Science methods for Multi-Messenger Astrophysics and Cosmology", as well as partial support from the Fondazione ICSC, Spoke 3 ``Astrophysics and Cosmos Observations'', Piano Nazionale di Ripresa e Resilienza Project ID CN00000013 ``Italian Research Center on High-Performance Computing, Big Data and Quantum Computing'' funded by MUR Missione 4 Componente 2 Investimento 1.4: Potenziamento strutture di ricerca e creazione di ``campioni nazionali di R\&S (M4C2-19 )'' - Next Generation EU (NGEU).
S.M. has received funding from the European Union’s Horizon 2020 research and innovation programme under the Marie Skłodowska-Curie grant agreement No. 101067043.
Y.Q. is supported by the Australian Research Council Centre of Excellence for All Sky Astrophysics in 3 Dimensions (ASTRO 3D), through project \# CE170100013.
R.T. acknowledges co-funding from Next Generation EU, in the context of the National Recovery and Resilience Plan, Investment PE1 – Project FAIR ``Future Artificial Intelligence Research''. This resource was co-financed by the Next Generation EU [DM 1555 del 11.10.22].
\section*{Data Availability}

The trained emulator is on a publicly accessible github repository, as well as available for install as a Python package using \texttt{pip}.



\bibliographystyle{mnras}
\bibliography{refs}




\appendix

\section{Parameter space dependence of the 21-cm PS emulation error}\label{sec:worst}
\begin{figure*}
    \centering
    \includegraphics[width =  \textwidth]{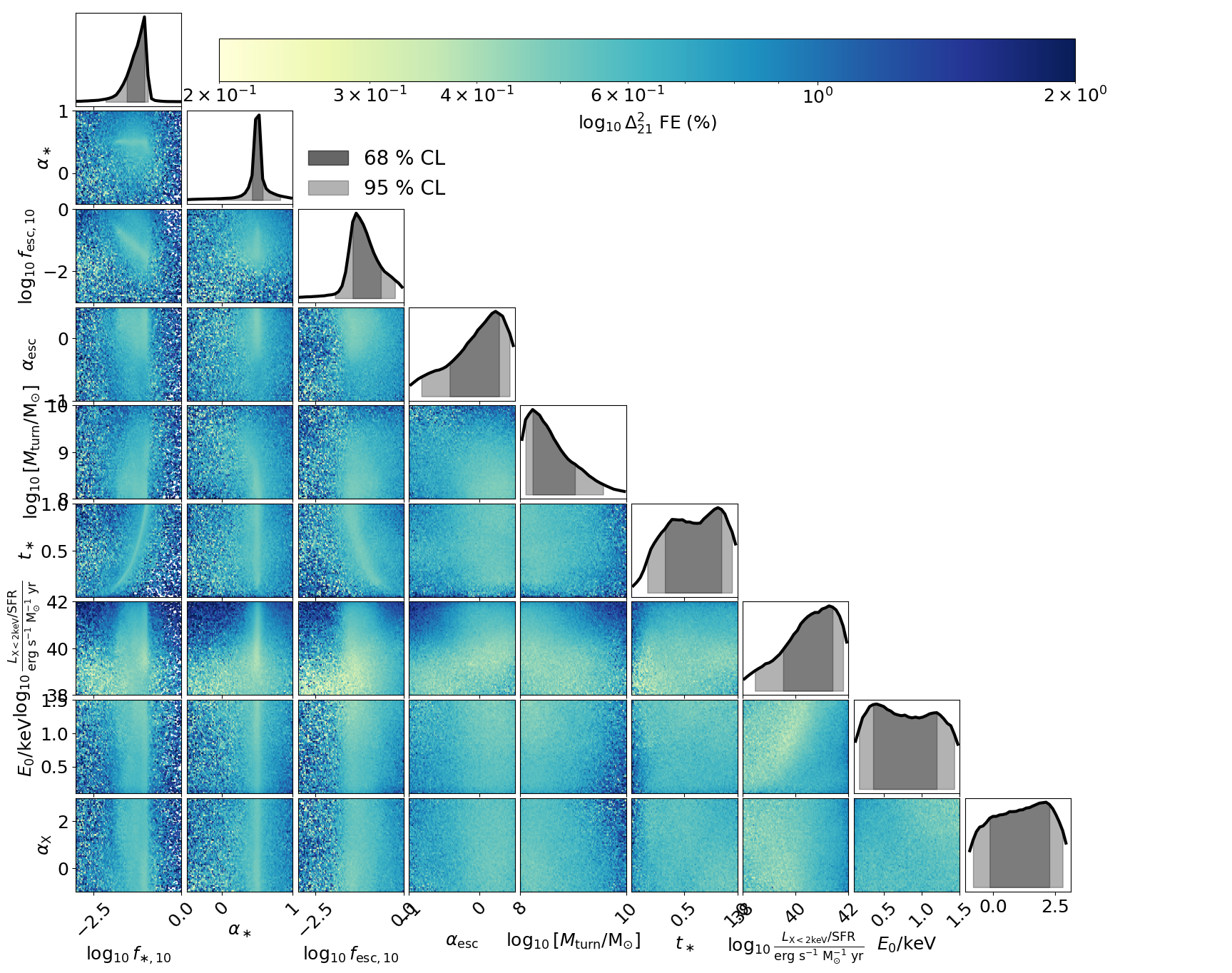}
    \caption{Distribution of the mean fractional error of the emulated $\log \Delta^2_{21}$. The colour of each bin in the 2D histogram is a weighted mean of the fractional error of the samples therein. On the diagonal, we show the 1D marginal density distribution of each astrophysical parameter in the test set. Note that the range of astrophysical parameters in the corner plot corresponds to the ranges taken for the flat prior of the inference used to generate the database.}
    \label{fig:PS_fe_corner}
\end{figure*}

In this Appendix, we look at how the emulation error is distributed over the 9D input parameter space. In Figure \ref{fig:PS_fe_corner}, we show the 21-cm power spectrum test set fractional error as a 2D histogram as a function of each pair of input astrophysical parameters. On the diagonal, we show the histogram (probability density) of each astrophysical parameter in the test set.
 
As expected, the emulation error peaks at the edges of parameter space where the density of samples is the lowest (see also Fig. 9 in \citealt{emuPSKern17} and top plot in Fig. 18 in \citealt{HERAYuxiang}).
However, the inclusion of the rejected livepoints in the training allowed our emulator to generalize well beyond the peak of the posterior (c.f. Fig. \ref{fig:corner_compare}).  Importantly, the mean FE remains modest ($\lesssim$ 2\%) throughout the prior volume.

\section{21cmSense Sensitivity Estimates for HERA's 6th-Season}
\label{app:sense}

To obtain mock error estimates for the forecasted sixth season of HERA observations, we used the updated open-source \texttt{21cmSense}\footnote{\url{https://github.com/rasg-affiliates/21cmSense}.} tool. The general algorithm of 21cmSense can be found in \cite{Pober14} and in the extensive documentation and tutorials of the updated codebase\footnote{e.g., \url{https://21cmsense.readthedocs.io/en/latest/tutorials/understanding_21cmsense.html}} (see also \cite{Liu2020} for a review including a similar argument). 
A brief outline of the calculations is as follows: \texttt{21cmSense} estimates thermal noise on any 3D  $\vec{k}$-mode as 
\begin{equation}
    P_N(\vec{k}_\perp, k_{||}) \propto \frac{T_{\rm sys}^2}{N_{k_{\perp}}\Delta \nu \tau_{\rm int}} \xi(\vec{k}_{\perp}, k_{||}),
\end{equation}
where $T_{\rm sys}$ is frequency-dependent system temperature
\begin{equation}
    T_{\rm sys} = T_{\rm sky}(\nu) + T_{\rm rcv}(\nu),
\end{equation}
$\Delta \nu = 122.07$\,kHz is the channel width of the observation and $\tau_{\rm int} = 300$\,sec is the coherently-averaged LST-bin size used in the analysis\footnote{In general, \texttt{21cmSense} uses the more fundamental snapshot integration time of the instrument, and re-phases observations over a longer `coherent observation duration', however HERA is a drift-scan telescope that performs no re-phasing, and all observations within an LST bin are considered coherent without re-phasing.}. 
Furthermore, $\xi$ is a `flag' function that takes the value 0 or 1 depending on the location of the 3D mode with respect to the foreground wedge (see below).

\begin{figure}
    \centering
    \includegraphics[width=\linewidth]{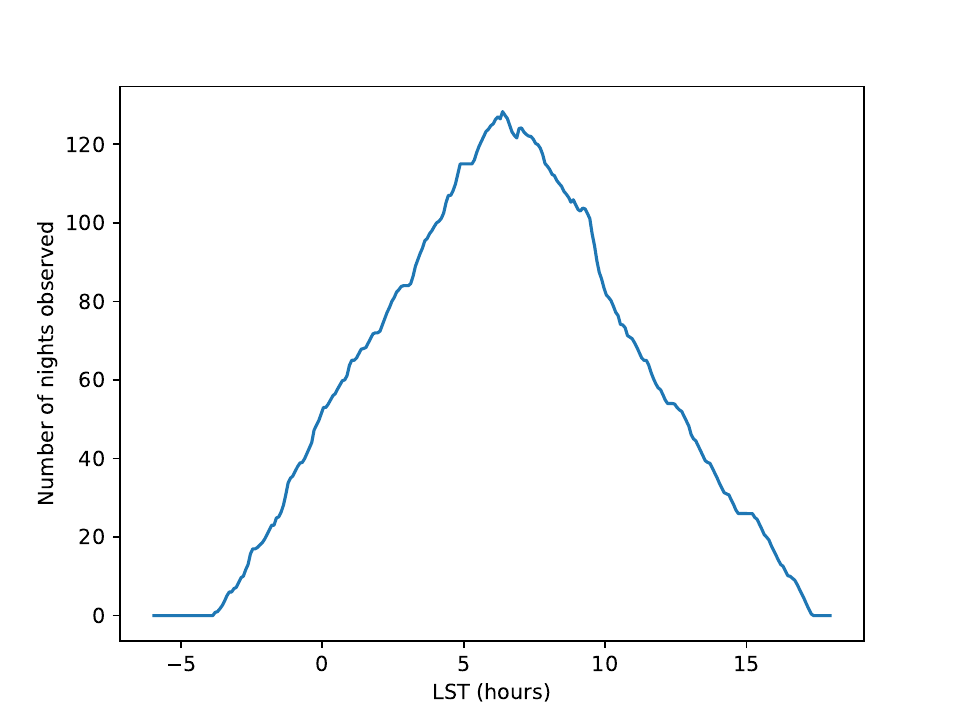}
    \caption{The number of times each 300-second LST bin was observed and un-flagged in HERA's sixth season, used for sensitivity estimates. Note that this accounts only for flags arising from strong effects that affect large swathes of the observed antennas and/or channels (eg. lightning storms, correlator outages), and further flags are applied in the downstream analysis.}
    \label{fig:nsamples_lstbin}
\end{figure}

In this equation, $N_{\rm k_\perp}$ represents the number of samples of this angular scale observed \textit{coherently} throughout the observing season (i.e., observations that are averaged together as visibilities).
In \texttt{21cmSense}, this is estimated by creating a grid on the $\vec{k}_\perp$ plane, whose cells are the size of the primary beam of the instrument in Fourier-space (for HERA, this is 7$\lambda$ at 150 MHz), and binning the the baseline coordinates into this grid\footnote{This is probably the greatest departure from the actual HERA analysis, which coherently averages only \textit{redundant} baselines, i.e., those that are equivalent to within several centimeters.}. 
In addition to the number of samples in a bin coming from different (redundant) baselines, we also have samples from the same baseline at different \textit{times}. Here, samples at the same LST on different nights are averaged coherently, but samples at different LSTs are averaged \textit{incoherently} (i.e., after forming power spectra). 
Currently, \texttt{21cmSense} only has support for specifying the number of nights observed and the number of hours observed each night (thereby specifying the number of LST bins in conjunction with the LST bin duration). However, in realistic observational programs, the same LST bins are not observed each night (whether due to the evolution of the sky throughout the season, or through flagging / data quality concerns). 
To partially account for this, we define a function $n_{\rm obs}({\rm LST})$ which counts the number of unflagged days observed over the season for any given 300-second-long LST bin (note that this accounts for flags of the entire observation, due to things like poor weather or correlator malfunctions, but not antenna- or channel-specific flags). To map this non-constant function of LST bin onto the schema available in \texttt{21cmSense}, which assumes the same LST bins are observed each night, we set
\begin{equation}
    n_{\rm days, eff} = \sqrt{\frac{\sum^{n_{\rm LST}}_{\rm LST}n^2_{\rm obs}({\rm LST})}{n_{\rm LST}}}
\end{equation}
and $t_{\rm day} = n_{\rm LST}\times 300\,{\rm sec}$. 
This achieves the same resulting thermal noise level, under the assumption that the sky temperature is constant over the LST bins. We use actual sixth-season HERA measurements for $n_{\rm obs}$, as shown in Fig. \ref{fig:nsamples_lstbin}.
We calculate $n_{\rm days, eff} = 67.4$ for coherent averaging and $t_{\rm day} = 21\,hr$ for incoherent averaging (i.e., the thermal noise from our observing pattern is equivalent to observing 253 300-second LST bins each for 67.4 days).
Finally, we apply a further factor of $\epsilon=0.9$ to $n_{\rm days, eff}$ to broadly account for finer-scale flags applied during analysis that are unaccounted in the LST-bin observing pattern of Fig. \ref{fig:nsamples_lstbin}.
In summary, we have
\begin{equation}
    N_{k_\perp} = \epsilon N_{{\rm bl}, k_{\perp}} N_{\rm days, eff} \sqrt{n_{\rm LST}}.
\end{equation}

The line-of-sight modes observed depend on the channel width, as already defined, and also the bandwidth of the observation. While HERA Phase II observes 200 MHz of bandwidth from 50 - 250\,MHz, power spectra are estimated in smaller `spectral windows' whose size is determined by a number of factors. 
Chiefly, the windows are as wide as possible, so as to include the largest scales where the signal is strongest, but are constrained by lightcone evolution \citep{Trott16,Datta12,21cmMC18} to be effectively smaller than $\sim10\,$MHz. In practice, spectral windows are chosen to lie between strongly flagged channels (eg. FM band and Orbcomm), which means their width is not constant. Here, we use constant 20\,MHz spectral windows, where we assume a Blackman tapering function is applied to each window to reduce the effective bandwidth to $\sim10\,$MHz (and an appropriate scaling factor of 1.737 is applied to the final noise level). We calculate noise estimates for all spectral windows between 50-250\,MHz, excluding the FM band between 90-110\,MHz.

We use a model for $T_{\rm sky}$ that is a power-law in frequency, with amplitude and spectral-index obtained from simulated auto-correlations of the diffuse sky, using the GSM \citep{Oliveira2008} and the HERA Phase II primary beam \citep{Fagnoni21} at LST=7\,hours:
\begin{equation}
    T_{\rm sky} = 150\, {\rm K} \times \left(\frac{\nu}{150\,{\rm MHz}}\right)^{-2.5}.
\end{equation}
Currently, \texttt{21cmSense} is not able to use different sky models for different LST-bins, so this choice represents the temperature for the most-observed LST bin. 
For $T_{\rm rcv}$, we use a frequency-dependent model based on electromagnetic simulations performed in \cite{Fagnoni21}, interpolated by a cubic spline. This model is close to a power-law at low frequencies, with an amplitude of $\sim600$\,K at 50\,MHz and asymptoting to a const $\sim60\,$K by 200\,MHz.

We construct several estimates of the noise based on different effective observing arrays. The sixth season of HERA data observed with a maximum of 209 antennas simultaneously in any given night (of the total 350 antennas available). The bulk of these antennas observed consistently throughout the season, though a fraction of them were swapped in and out. In our estimates here, we assume that the same antennas observe consistently throughout the season, which is a reasonable approximation. 
Nevertheless, in practice, even though 209 antennas are being correlated at any given moment, some fraction of them are flagged over all channels (e.g., due to swapped polarizations, non-redundancies from physical effects such as feed displacement, or X-engine failures that affect a subset of antennas, etc.). 
The average number of antennas actually observing per-night throughout the season is as-yet unknown, though initial estimates place it at $\sim150$ antennas \citep{HERAMemo122}. Here we use $N_{\rm ants} = 148$, where the antennas are drawn randomly from the set of 209 antennas that actually observed throughout the season.
In all cases, we use only baselines whose East-West length is greater than 15\,m (i.e., we exclude North-South baselines, as their systematics are more difficult to filter out), and also only include baselines shorter than 150\,m, similar to analyses of previous HERA seasons. 

After obtaining the 3D sensitivity grid, we incoherently average into 1D spherical $|k|$-shells with bins of width $\Delta k_{||}$. 
In this process, we flag out $(|k_\perp|, k_{||})$-bins within the foreground `wedge' \citep{Liu14I, Liu14II}, defined by 
\begin{equation}
    k^{\rm wedge}_{||} = 0.15 h{\rm Mpc}^{-1} + \frac{{\rm d} k_{||}}{{\rm d}\eta}(\nu) \frac{|b|}{c},
\end{equation}
with $|b|$ the baseline length (in meters) corresponding to a given $k_{\perp}$, and $dk_{||}/d\eta$ a redshift-dependent cosmological factor converting bandwidth into cosmic distance.
This corresponds to the `horizon' limit of foregrounds in delay-space, plus a conservative buffer of 0.1 $h$/Mpc (corresponding to the buffer used in first-season HERA analyses).

In addition to the thermal variance, cosmic- (or sample-) variance is added, proportional to a fiducial cosmological power spectrum, $P^2_{\rm theory}$ divided by the number of LST bins and $k_\perp$-modes in a spherical shell. We note that using the number of LST-bins is inspired by the idea that LST-bins should capture the entire duration of `coherence', equal to roughly the beam-crossing time for an antenna. However, HERA is conservative in using shorter coherence times, resulting in many more LST-bins. This reduces the thermal sensitivity, but artificially reduces the cosmic variance estimated by \texttt{21cmSense}. Nevertheless, since cosmic variance is generally a sub-dominant contribution to the total variance, this should not have a large effect on the results presented here. For the fiducial theoretical model, we here use the model from \cite{EOS2021}.

\begin{figure*}
    \centering
    \includegraphics[width =  0.8\textwidth]{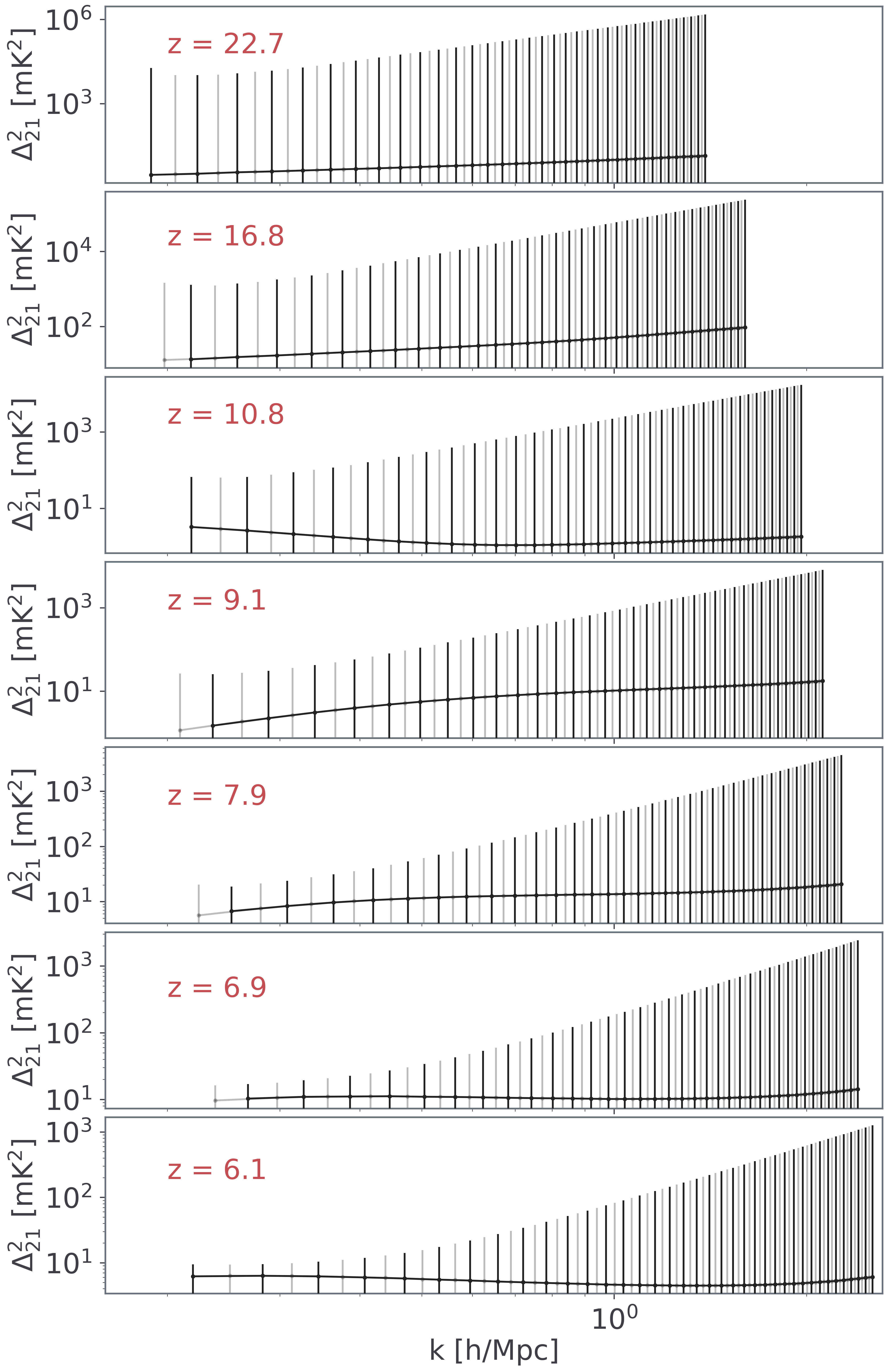}
    \caption{HERA phase II 6th season sensitivity forecast obtained using \texttt{21cmSense} with the parameters specified in Table \ref{tab:sense}. Note that in practice, HERA decimates the $k$-bins to avoid requiring non-diagonal covariance (e.g., \citealt{HERAYuxiang}).  Here we have approximated this practice by using only half of the above $k$-bins (those highlighted in black) when computing the likelihood for our inference.}
    \label{fig:H6C_sens}
\end{figure*}

We summarize the parameters used in Table \ref{tab:sense} and show the full HERA phase II 6th season sensitivity forecast in Figure \ref{fig:H6C_sens}.

\begin{table*}
    \centering
    \begin{tabular}{c|c|c}
         Parameter &  Description & Value \\
         \hline
         $N_{\rm ants}$ & Number of antennas within the 209 available actually observed. & 75, 100, 120, 148, 209 \\
         $T_{\rm sky}$ & Sky temperature model & $150 {\rm K} (\nu/150 {\rm MHz})^{-2.5}$ \\
         $T_{\rm rcv}$ & Receiver Temperature & Empirical, 600 K at 50 MHz, 60 K above 200 MHz \\
         $\Delta \nu$ & Channel width & 122.07 kHz \\
         $\tau_{\rm int}$ & Coherent integration time (LST bin width) & 300 sec \\
         $\Delta$ u & UV-grid size for coherent baseline averaging & 7 $\lambda$ \\
         $N_{\rm days, eff}$ & Effective number of days observed coherently & 67.4$^\dagger$ \\
         $t_{\rm day}$ & Effective observed hours per day & 21 hours (253 LST bins) \\
         $\epsilon$ & Efficiency factor for frequency-dependent flags & 0.9 \\
         $B$ & Spectral window bandwidth & 20 MHz \\
         $B_{\rm eff}$ & Effective spectral window bandwidth after Blackman taper & 11.51 MHz \\
         FG Wedge Level & Line-of-sight scale below which modes are filtered & 0.15 h/Mpc + horizon \\
         Theory Model & Cosmological power spectrum from which to calculate cosmic variance & \citealt{EOS2021} \\
         \hline
    \end{tabular}
    \caption{Parameters used within \texttt{21cmSense} to obtain sensitivity estimates for the sixth season of HERA observations. See App. \ref{app:sense} for details on the algorithm. $^\dagger$  note that $N_{\rm days, eff}$ and $t_{\rm int}$ are effectively equivalent to the actual LST footprint of the season in terms of thermal noise, under the assumption that the sky temperature is constant with LST.}
    \label{tab:sense}
\end{table*}

There are a few caveats to these estimates. Most importantly, baselines found within 7$\lambda$ UV-bins together are considered redundant, while in the HERA analysis only baselines within 10\,cm are considered redundant. This will artificially increase thermal sensitivity estimates. Secondly, the sky temperature is considered constant over the LST bins. To minimize the effect of this limitation, we use a sky model that is based at the most-observed LST (7 hours). Thirdly, cosmic variance is reduced as the square root of the number of LST bins, instead of the number of independent `fields' observed. This artificially increases the sensitivity from cosmic variance, though this should not have a large effect since this is the sub-dominant contribution on most scales and redshifts.
Finally, in this forecast we did not decimate the $k$-bins as was done in previous analyses. 
This results in some unaccounted covariance between $k$-bins that would tend to over-estimate the sensitivity. We do not expect this to significantly affect the qualitative conclusions derived from the forecast.

\bsp	
\label{lastpage}
\end{document}